\documentclass[preprint2]{aastex}

\usepackage{lineno}

\slugcomment{Not to appear in Nonlearned J., 45.}

\begin{document}


\title{Spectral Lags and the Lag-Luminosity Relation: An Investigation
with $Swift$ BAT Gamma-ray Bursts}



\author{
T. N. Ukwatta \altaffilmark{1,2}, M. Stamatikos
\altaffilmark{2,3}, K. S. Dhuga \altaffilmark{1}, T. Sakamoto
\altaffilmark{2,4,5}, S. D. Barthelmy \altaffilmark{2}, A.
Eskandarian \altaffilmark{1}, N. Gehrels \altaffilmark{2}, L. C.
Maximon \altaffilmark{1}, J. P. Norris \altaffilmark{6} and W. C.
Parke \altaffilmark{1}}


\altaffiltext{1}{The George Washington University, Washington,
D.C. 20052.}

\altaffiltext{2}{NASA Goddard Space Flight Center, Greenbelt, MD
20771.}

\altaffiltext{3}{Center for Cosmology and Astro-Particle Physics
(CCAPP) Fellow/Department of Physics, The Ohio State University,
191 West Woodruff Avenue, Columbus, OH 43210.}

\altaffiltext{4}{Center for Research and Exploration in Space
Science and Technology (CRESST), NASA Goddard Space Flight Center,
Greenbelt, MD 20771.}

\altaffiltext{5}{The University of Maryland, Baltimore County,
Baltimore, MD 21250.}

\altaffiltext{6}{University of Denver, Department of Physics and
Astronomy, 2112 East Wesley Ave. Room 211, Denver CO 80208}


\begin{abstract}
Spectral lag, the time difference between the arrival of
high-energy and low-energy photons, is a common feature in
Gamma-ray Bursts (GRBs). Norris et al. 2000 reported a correlation
between the spectral lag and the isotropic peak luminosity of GRBs
based on a limited sample. More recently, a number of authors have
provided further support for this correlation using arbitrary
energy bands of various instruments. In this paper we report on a
systematic extraction of spectral lags based on the largest
$Swift$ sample to date of 31 GRBs with measured redshifts. We
extracted the spectral lags for all combinations of the standard
$Swift$ hard x-ray energy bands: 15-25 keV, 25-50 keV, 50-100 keV
and 100-200 keV and plotted the time dilation corrected lag as a
function of isotropic peak luminosity. The mean value of the
correlation coefficient for various channel combinations is -0.68
with a chance probability of $\sim 0.7 \times 10^{-3}$. In
addition, the mean value of the power-law index is $1.4 \pm 0.3$.
Hence, our study lends support for the existence of a
lag-luminosity correlation, albeit with large scatter.
\end{abstract}






\section{Introduction}
After decades of research, a satisfactory explanation of the
temporal behavior of Gamma-ray Burst (GRB) light-curves is still
lacking. Despite the diversity of GRBs, some general
characteristics and correlations have been identified: Spectral
lag is one such characteristic. The spectral lag is the difference
in time of arrival of high-energy pulses verses low-energy pulses.
In our analysis a positive spectral lag corresponds to an earlier
arrival time for the higher energy photons. The observed spectral
lag is a common feature in
GRBs~\citep{cheng1995,norris1996,band1997}. The study of spectral
lag between energy bands, which combines temporal and spectral
information, potentially can constrain GRB
models~\citep{Lu2006,Shen2005,Qin2004,Schaefer2004,Ioka2001,Salmonson2000}.

Based on six GRBs with known redshifts, \cite{norris2000} found an
anti-correlation between the spectral lag and the isotropic peak
luminosity. Further evidence for this correlation was provided by
\cite{norris2002}, \cite{gehrels2006}, \cite{Schaefer2007},
\cite{Stamatikos2008} and \cite{Hakkila2008}. Others have used
this relation as a redshift indicator~\citep{Murakami2003,
band2004} and as a cosmological tool~\citep{2003ApJ...594..674B,
Schaefer2007, 2008ApJ...685..354L, 2008A&A...487...47M}.

\cite{Hakkila2008} have used a pulse-profile fitting technique (a
four-parameter pulse function introduced by \cite{norris2005}) to
show that the correlation is between lags of the pulses and the
luminosity of the pulses seen in GRBs. However, the method is
limited because it applies only to very bright bursts where pulses
are clearly identifiable and described by the assumed pulse
profile.

Many authors have tried to explain the physical cause of the
lag-luminosity relation and a number of models have been proposed.
\cite{Salmonson2000} argues that the anti-correlation is due to
the variations in the line-of-sight velocity of various GRBs.
\cite{Ioka2001} suggest that the relation is a result of
variations of the off-axis angle when viewing a narrow jet.
\cite{Schaefer2004} invokes a rapid radiation cooling effect to
explain the correlation. This effect tends to produce short
spectral lags for highly luminous GRBs.

Regardless of its physical origin, the spectral lag is an
important measurement for GRB science because of its usefulness in
differentiating long and short GRBs ~\citep{Kouveliotou1993}: Long
bursts give large lags and short bursts give relatively small
lags~\citep{norris1995, norris2006}. Even though a few exceptions
to this classification scheme have been found, such as GRB
060614~\citep{gehrels2006}, the GRB community still continues to
use the spectral lag as one of the classification criteria. Note
that more elaborate classification schemes based on multiple
observational parameters, such as the host galaxy property, has
also been proposed \citep{Donaghy2006,Zhang2009}.

Moreover, based on the analysis of GRB 080319B,
\cite{Stamatikos2009} show that there is a possible correlation
between the prompt optical emission and the evolution of spectral
lag with time.

Most of the previous work on spectral lags has been based on
observations with the Burst and Transient Source Experiment
(BATSE) on the Compton Gamma Ray
Observatory~\citep{tsutsui2008,Hakkila2008,Hakkila2007,Chen2005,band2004,Salmonson2002,norris2002,band1997}.
The launch of the $Swift$ satellite~\citep{gehrels2004} ushered in
a new era of GRB research. In this paper we present a detailed
study of spectral lags using a subset of Swift Burst Alert
Telescope (BAT) data.

The structure of the paper is the following: In
section~\ref{methodology} we discuss our methodology with a case
study featuring GRB 060206. In section~\ref{results} we present
our results for a sample of 31 $Swift$ BAT long bursts and
investigate the lag-luminosity relation for various channel
combinations. Finally, in section~\ref{discussion} we discuss some
implications of our results. Throughout this paper, the quoted
uncertainties are at the 68\% confidence level.

\section{Methodology} \label{methodology}
\subsection{Light Curve Extraction}

$Swift$ BAT is a highly sensitive instrument, which utilizes a
coded aperture mask to localize bursts \citep{barthelmy2005}. The
basic imaging scheme is that a gamma-ray source illuminates the
coded aperture mask, and casts a shadow onto a position sensitive
detector. Each position in the sky will produce a unique shadow
pattern in the detector plane. Hence by comparing the observed
shadow with precalculated shadow patterns for all possible points
in the sky it is possible to find the actual position of the
source that created the given shadow pattern. However, in practice
each detector can be illuminated by many points on the sky whereas
each point on the sky can illuminate many detectors. To
disentangle each point in the sky, special software designed by
the $Swift$ BAT team is used.

In order to generate light curves, a process called mask weighting
is utilized. The mask weighting assigns a ray-traced shadow value
for each individual event, which then enables the user to
calculate light curves or spectra. We used the
\texttt{batmaskwtevt} and \texttt{batbinevt} tasks in FTOOLS to
generate mask weighted, background-subtracted light curves for our
analysis. Resulting light curves and their uncertainties are
calculated by propagation of errors from raw counts (subject to
Poissonian noise).

\subsection{The Cross Correlation Function and Spectral Lag}
\label{CCF_cal}

There are at least three well known ways of extracting spectral
lags; (1) pulse peak-fit method~\citep{norris2005,Hakkila2008},
(2) Fourier analysis method~\citep{Li2004}, and (3)
cross-correlation function (CCF) analysis method
~\citep{cheng1995,band1997}. The pulse peak-fit method gives a
simple straight forward way for extracting lags. It does however
assume a certain pulse function for the pulses in the light curve
and may also be limited to very bright bursts. It is not
immediately clear how this method would fare in cases where the
light curves are sufficiently complex i.e., not dominated by a
prominent pulse. For transient events such as GRBs using the
Fourier analysis technique also has its
difficulties~\citep{Li2004}. Since GRB light curves do not exhibit
obvious periodicities, Fourier transforms typically yield a large
number of coefficients to describe their temporal structure. These
coefficients, in turn, produce a spectral lag value for each
corresponding frequency component i.e., a spectrum of lags is
generated. The generated spectra exhibit a variety of shapes
depending on the complexity of the light curve~\citep{Li2004} thus
making the extraction of an intrinsic lag questionable. Hence, in
this work, we develop a method to calculate the time-averaged
spectral lag and its uncertainty via a modification of the CCF
method.

The use of the Pearson cross-correlation function is a standard
method of estimating the degree to which two series are
correlated. For two counting series $x_i$ and $y_i$ where
$i=0,1,2,...(N-1)$, the CCF with a delay $d$ is defined as
\begin{equation}\label{eq:no1}
CCF_{\tiny \textrm{Std}}(d, x, y)=\frac{\sum_{i=1}^{N-d}(x_i- \bar
x)(y_{i+d}-\bar y)}{\sqrt{\sum_{i}(x_i-\bar
x)^2}\sqrt{\sum_{i}(y_{i}-\bar y)^2}}.
\end{equation}
Here $\bar x$ and $\bar y$ are average counts of the two series
$x$ and $y$ respectively.
The denominator in the expression above serves to normalize the
correlation coefficient such that $-1 \leq CCF_{\tiny
\textrm{Std}}(d, x, y) \leq 1$, the bounds indicating maximum
correlation and zero indicating no correlation. A high negative
correlation indicates a high correlation but of the inverse of one
of the series. Note that the time delay ($\tau$) is given by $\tau
= d\, \times$ time bin size.

However, \cite{band1997} proposed that for transient events such
as GRBs, non-mean subtracted definition given below is more
suitable for the time-averaged lag.
\begin{equation}\label{eq:no2}
CCF_{\tiny \textrm{Band}}(d, x, y)=\frac{\sum_{\small
i=\textrm{max}(1, 1-d)}^{\small \textrm{min}(N, N-d)}x_i \,
y_{i+d}}{\sqrt{\sum_{i}x_i^2 \, \sum_{i}y_{i}^2}}
\end{equation}
We have tested both definitions of the CCF using synthetic light
curves with artificially introduced spectral lags. Our tests
showed that the $\rm CCF_{\tiny \textrm{Band}}$ consistently
recovered the introduced lag while $\rm CCF_{\tiny \textrm{Std}}$
sometimes failed (possible reasons for this failure are noted in
\cite{band1997}). Hence in our analysis we used the $\rm
CCF_{\tiny \textrm{Band}}$ definition and from this point onwards
in the paper we refer to it simply as the CCF.

For a given pair of real light curves, we determine the CCF using
Equation \ref{eq:no2}. At this stage the resulting CCF values do
not have any uncertainties associated with them. In order to
determine these uncertainties, we use a Monte Carlo simulation.
Here we make 1,000 Monte Carlo realizations of the real light
curve-pair based on their error bars as shown below.
\begin{equation}\label{eq:no3}
\rm LC^{simulated}_{bin} = LC^{real}_{bin} + \xi \times LC^{real
\,error}_{bin}
\end{equation}
Where $\xi$ is a random number generated from a gaussian
distribution with the mean equal to zero and the standard
deviation equal to one. For each simulated light curve-pair we
calculate the CCF value for a series of time delays. This results
in a 1,000 CCF values per time-delay bin. The standard deviation
of these values per time-delay bin is then assigned as the
uncertainty in the original CCF values obtained from the real
light curves.

\subsection{Extracting Spectral Lags} \label{lag_extraction}

We realize there may be number of ways to define the spectral lag,
but in this work, we define it as the time delay corresponding to
the global maximum of the cross-correlation function. To locate
this global maximum, we fit a Gaussian curve to the CCF. The
uncertainties in the CCF are obtained using a Monte Carlo
procedure discussed in section \ref{CCF_cal}. In essence, our
fitting procedure locates the centroid of the cross correlation
function and is thus relatively insensitive to spurious spikes in
the CCF. We tested and verified the robustness of this procedure
by performing a number of simulations in which artificial lags
were first introduced into the light curves and then successfully
recovered. In addition, our tests with these artificial light
curves show that the CCF can become asymmetric (around its global
maximum) if the shape of one of the light curves is significantly
different from the other. This energy dependent feature
potentially requires a more complex fitting function than a
Gaussian or a quadratic to fit the CCF over the entire range.
Instead of resorting to a more complex fitting function we were
able to recover the (known) lags by fitting the CCFs (with a
Gaussian) over limited but asymmetric ranges.

\subsubsection{Time Bin Selection}

For $Swift$ GRBs the minimum time binning is 0.1 ms but one can
arbitrarily increase this all the way up to the duration of the
burst. It is important to understand the effect of time binning on
the extracted spectral lags. Presumably, by changing the time
binning of the light curve one is affecting the signal-to-noise
ratio. By employing increasingly coarser binning one is averaging
over the high frequency components of the light curve. Clearly,
one has to be careful not to use overly large time bin sizes
otherwise one risks losing the sought-after information from the
light curve.

\begin{figure}
\epsscale{1.0} \plotone{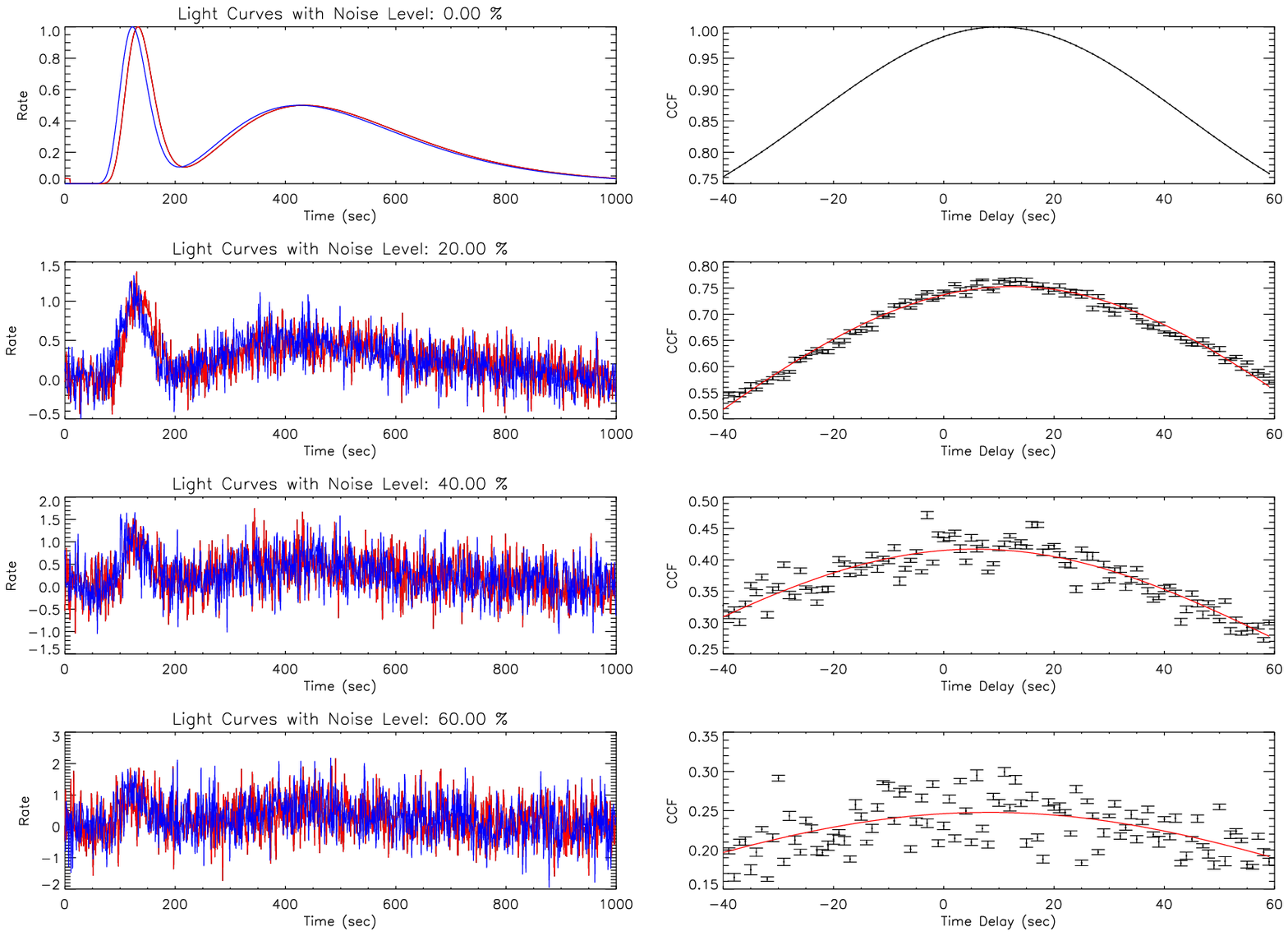} \caption{The effect of noise
on the CCF. Panels on the left show two synthetic light curves, in
which a 10-second artificial lag is added. From top to bottom the
noise level is increased as 0\%, 20\%, 40\% and 60\% respectively.
The corresponding CCF vs time delay plots are shown in the right
panels along with Gaussian fits.}\label{LC_CCF}
\end{figure}

In order to understand the effect of time binning more fully, we
did a number of simulations utilizing peak normalized synthetic
light curves (composed of FRED\footnote{Fast Rise Exponential
Decay}-like pulse shapes with Gaussian distributed noise) in which
artificial lags were introduced. We incrementally increased the
noise level and studied its effect on the maximum correlation
value in the CCF vs time delay (CCFMax) plot. In Fig. \ref{LC_CCF}
we display the synthetic light curves with several noise levels
(0\%, 20\%, 40\% and 60\% respectively) as well as the calculated
CCF with typical Gaussian fits. As expected, the CCFMax value (see
the right panel of Fig.~\ref{LC_CCF}) decreases gradually as the
noise level increases. We also note that the scatter in the CCF
increases considerably with the noise level. The global maximum in
the CCF is clearly visible at the 40\% noise level and a good fit
is obtained with a Gaussian. However, this is not the case for the
60\% noise-level curve, in which the scatter is quite significant,
and the CCF global maximum is barely visible leading to a poor
fit.

\begin{figure}
\epsscale{1.0} \plotone{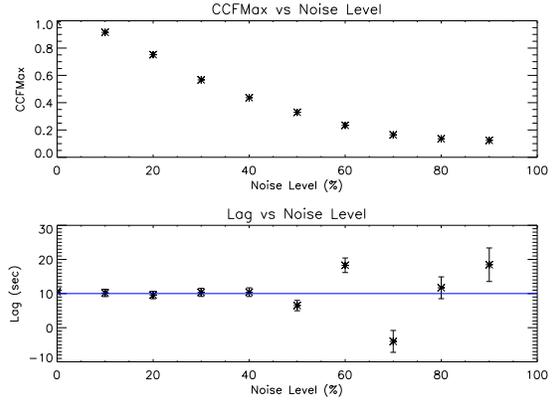} \caption{The
effect of noise on the maximum correlation of the CCF (CCFMax) and
the extracted spectral lag. The horizontal line (blue) in the
bottom panel indicates the 10-second artificial
lag.}\label{CCFMaxVsNoiseLevel}
\end{figure}

In Fig.~\ref{CCFMaxVsNoiseLevel} we show the behavior of the
CCFMax value and the extracted spectral lag as a function of the
noise level. We first note that the CCFMax value smoothly tracks
the signal-to-noise level in the light curves (see the upper panel
of Fig.~\ref{CCFMaxVsNoiseLevel}). Secondly, we note that the
extracted lag value agrees well with the artificially introduced
lag of 10 seconds up to a noise level of about 40\%. We further
note that the scatter in the extracted lag value increases as the
noise contribution increases beyond 40\%. Although it is not
immediately obvious from this figure, CCF values above a noise
level of 40\% show large scatter (see bottom right panel of
Fig.~\ref{LC_CCF}) thus making the extracted lag value uncertain.
This is directly reflected by the increasing error bars in the
extracted value. These simulations were repeated for number of
time lags and in all cases similar results were obtained, in
particular, the behavior of the CCFMax as a function of the noise
level was confirmed. Based on the results of these simulations, we
chose a CCFMax $\sim 0.5$, corresponding to a noise level of about
40\%, as our guide for picking the appropriate time binning.

Procedurally, we start with a time bin size of 1024 ms and
decrease the time binning by powers of two until the CCFMax
becomes $\sim 0.5$ and use that time bin size as the preferred
time binning for the lag extraction. By using this procedure we
are able to arrive at a reasonable bin size that preserves the
fine structure in the light curve and at the same time keeps the
contribution of the noise component at a manageable level.

\subsubsection{Uncertainty in Spectral Lags} \label{lag_error}

We have studied three methods to determine the uncertainty in the
extracted spectral lags. The first method is to use the
uncertainty that is obtained by fitting the CCF with a Gaussian
curve. The second method is an adaptation of equation (4) used in
\cite{Gaskell1987}
\begin{equation}\label{equation4}
\sigma_{\rm \, lag} = \frac{0.75 \, W_{\rm HWHM}}{1+h\sqrt{n-2}}.
\end{equation}
Here $W_{\rm HWHM}$ is the half-width half-maximum of the fitted
Gaussian, $h$ the maximum height of the Gaussian and $n$ is the
number of bins in the CCF vs time delay plot. This method utilizes
more information about the fit and the CCF such as the width,
height, and number of bins to estimate the uncertainty. The third
method utilizes a Monte Carlo simulation. We found that the first
method gives systematically smaller uncertainty in the lag by a
factor of two or more relative to the other two methods. The
second and third methods give comparable values. We adopted the
most conservative of the three methods (i.e. the one based on the
Monte Carlo simulation) to determine the uncertainties in the lag.

\subsubsection{Lag Extraction: Case Study}

\begin{figure}
\epsscale{1.0} \plotone{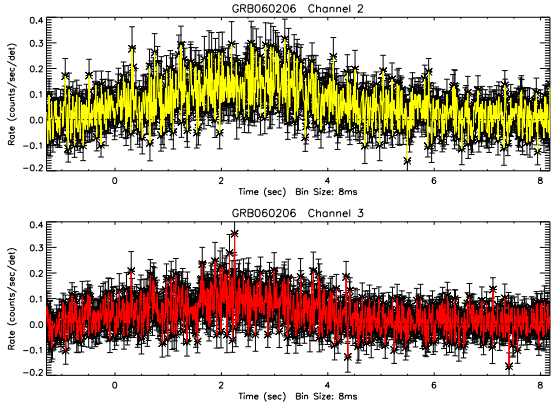}
\caption{Swift-BAT prompt gamma-ray (8 ms time bin) light curves
for GRB 060206 with canonical energy channels 2 (25-50 keV) and 3
(50-100 keV).}\label{LC}
\end{figure}

\begin{figure}
\epsscale{1.0} \plotone{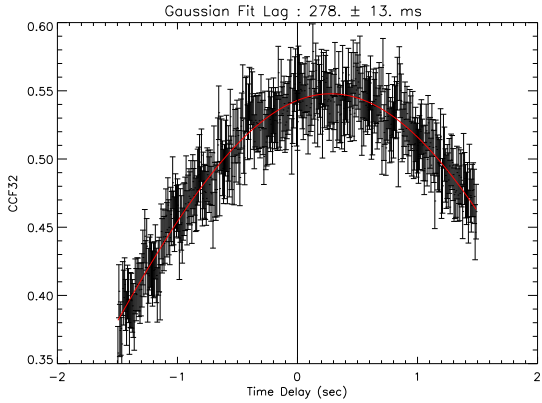} \caption{CCF as a
function of time delay for the two light curves in Fig.\ref{LC}.
The time delay corresponding to the peak of the Gaussian fit is
the spectral lag of the burst, which is 278 $\pm$ 13 ms. The
uncertainty quoted here is from the fit, which tends to be factor
of two or more less than the value obtained by the Monte Carlo
simulation shown in Fig.~\ref{Sim}.}\label{CCF}
\end{figure}

\begin{figure}
\epsscale{1.0} \plotone{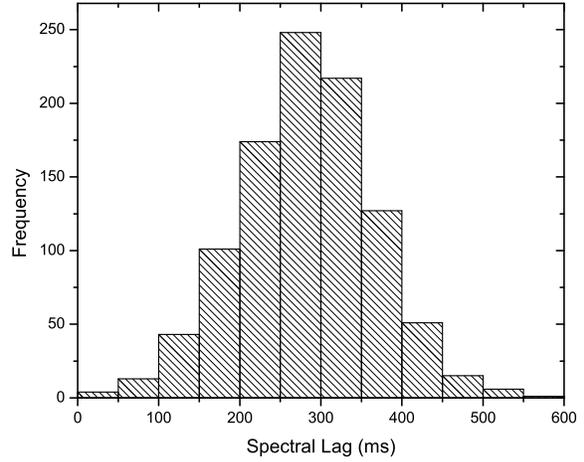}
\caption{Histogram of 1,000 simulated spectral lag values. We take
the standard deviation of the distribution of simulated spectral
lag values as the uncertainty of the fitted spectral lag value
which was found in Fig.~\ref{CCF}. The final spectral lag value is
278 $\pm$ 74 ms.}\label{Sim}
\end{figure}

To illustrate the lag extraction procedure more clearly, we
present a case study using GRB 060206. The light curve segment is
selected by scanning both forward and backward directions from the
peak location until the count rate drops to less than 5\% of the
peak count rate (using $15-200$ keV light curve). This selection
method is chosen to include the most intense segment of the burst
and to capture any additional overlapping pulses near the main
structure. Presumably, these pulses also contribute to the overall
spectral lag. In the case of GRB 060206 this corresponds to a
light curve segment starting 1.29 seconds prior to the trigger and
8.18 seconds after the trigger (see Fig.~\ref{LC}). Next we
calculate the CCF and plot it as a function of time delay as shown
Fig.~\ref{CCF}. Error bars on the CCF points were obtained via a
Monte Carlo simulation of 1,000 realizations of the original light
curves (see section \ref{CCF_cal}). As noted earlier, we start
with time bin size of 1024 ms and decrease the time binning by
powers of two until the CCFMax becomes $\sim 0.5$ for a given
channel combination, in this case BAT standard channel 2 ($25-50$
keV) and 3 ($50-100$ keV). For GRB060206 channel 2 and 3, the time
bin size correspond to 8 ms. The global maximum of the CCF vs time
delay plot corresponds to the spectral lag and its value is
obtained by fitting a Gaussian curve. We choose a range of the
time delay (in this case from -1.5 seconds to 1.5 seconds)
manually to identify the global maximum. In order to obtain the
uncertainty in the spectral lag, we employ another Monte Carlo
simulation, in which we create 1,000 additional realizations of
the input light curves as described in section \ref{CCF_cal}, and
repeat the previously described process for the simulated light
curves. A histogram of the resulting (1,000) spectral lag values
is shown in Fig.~\ref{Sim} for GRB 060206. The standard deviation
of these values is the uncertainty in the spectral lag.

\subsection{Isotropic Peak Luminosity}

To compare observations with different instruments we need to
calculate flux over some fixed energy band. In order to do this we
need to know the best--fit spectral function to the observed
spectrum and its spectral parameters. Often, GRB spectra can be
well fitted with the Band function~\citep{band1993}, an empirical
spectral model defined as follows:
\begin{displaymath}
N = \left\{ \begin{array}{ll}A
(\frac{E}{100\,\rm{kev}})^{\,\alpha}\,e^{-(2+\alpha)E/E_{\rm
p}},\,\small E\,\le\,\big(\frac{\alpha -
\beta}{2+\alpha}\big)E_{\rm p}& \\
A (\frac{E}{100\,\rm{kev}})^{\,\beta}\,[\frac{(\alpha-\beta)E_{\rm
p}}{(2+\alpha)100\,\rm{keV}}]^{\alpha-\beta}\,e^{(\beta-\alpha)},
\,\rm otherwise. &
\end{array} \right.
\end{displaymath}
There are four model parameters in the Band function; the
amplitude (A), the low-energy spectral index ($\alpha$), the
high-energy spectral index ($\beta$) and the peak of $\nu F_{\nu}$
spectrum ($E_{\rm p}$).

If the GRB spectrum is well described by the Band function, then
the values of $\alpha$, $\beta$, $E_p$ and the observed peak flux,
$f_{\rm{obs}}$, in a given energy band ($E_{\rm min}$ and $E_{\rm
max}$) are often reported. We can calculate the normalization $A$
with
\begin{equation}
A =\,\frac{f_{\rm obs}}{\int_{E_{\rm min}}^{E_{\rm max}}\,N'(E)\,
dE}
\end{equation}
where $f_{\rm obs}$ is given in $\rm photons \, cm^{-2}\,s^{-1}$
and $N'=N/A$.

The observed peak flux for the source-frame energy range $E_{1}
=1.0\,\rm{keV}$ to $E_{2} = 10,000\,\rm{keV}$ is
\begin{equation}
f_{\rm obs}^{\rm new} = \int_{E_{1}/(1+z)}^{E_{2}/(1+z)}\,N(E)E\,
dE.
\end{equation}
The isotropic peak luminosity is
\begin{equation}
L_{\rm iso}= 4 \pi d_L^{\,2} \, f_{\rm{obs}}^{\rm new}
\end{equation}
where $d_L$ is the luminosity distance given by,
\begin{equation}
d_L=\frac{(1+z)c}{H_0}\int_{0}^{z} \frac{dz'}{\sqrt{\Omega_M
(1+z')^3 + \Omega_L}}
\end{equation}
For the current universe we have assumed, $\Omega_M = 0.27$,
$\Omega_L = 0.73$ and the Hubble constant $H_0$ is
$70\,\rm(kms^{-1})/Mpc \,=\,2.268 \times 10^{-18}\,\rm
s^{-1}$~\citep{Komatsu2009}.

To determine the uncertainty in $L_{\rm iso}$, we employ a Monte
Carlo simulation. We simulate spectral parameters $\alpha$,
$\beta$, $E_{\rm p}$ and flux assuming their reported value as
sample mean and reported uncertainty as sample standard deviation,
then calculate $L_{\rm iso}$, for 1,000 variations in these
parameters. If a parameter has uneven uncertainty values then each
side around the parameter is simulated with different uncertainty
values as standard deviation. Then we take the 16th and the 84th
ranked values ($1 \sigma$ uncertainty) as the lower limit and the
upper limit of $L_{\rm iso}$ respectively.

\section{Results} \label{results}

We selected a sample of long GRBs ($T90 > 2$ sec, excluding short
bursts with extended emission), detected by $Swift$ BAT from 2004
December 19 to 2009 July 19, for which spectroscopically confirmed
redshifts were available. Out of this initial sample (102), a
subset of 41 GRBs were selected with peak rate $> 0.3$
counts/sec/det ($15-200$ keV, 256 ms time resolution). Finally, we
selected 31 GRBs for which a clear global maximum can be seen in
the CCF vs time delay plots with maximum correlation of at least
0.5 (with 256 ms time binning) for all channel combinations.
The spectral parameters of the final sample are given in
Table~\ref{tab:para}. We note that our final sample contains
bursts with redshifts ranging from 0.346 (GRB 061021) to 5.464
(GRB 060927) and the average redshift of the sample is $\sim$2.0.

Out of our sample, 18 bursts have all Band spectral parameters
measured and comprise our ``Gold'' sample. The remaining 13 bursts
are further divided into ``Silver'' and ``Bronze'' samples. In the
``Silver'' sample, 10 bursts have $E_{\rm p}$ determined by
fitting a cutoff power-law\footnote{$dN/dE \sim E^\alpha
\exp{(-(2+\alpha)E/E_{\rm p})}$} (CPL) to spectra and for GRB
060418, $E_{\rm p}$ is reported without uncertainty, so we assumed
a value of 10\%. These 10 bursts do not have the high-energy
spectral index, $\beta$, measured, so we used the mean value of
the BATSE $\beta$ distribution, which is $-2.36\pm0.31$
\citep{Kaneko2006, Sakamoto2009}. The ``Bronze'' sample
(consisting of 3 bursts) does not have a measured $E_{\rm p}$. We
have estimated it using the power-law index ($\Gamma$) of a simple
power-law (PL) fit as described in \cite{Sakamoto2009}.  For these
3 bursts, the low-energy spectral index, $\alpha$ and the
high-energy spectral index, $\beta$, were not known, so we used
the mean value of the BATSE $\alpha$ and $\beta$ distribution,
which is $-0.87\pm0.33$ and $-2.36\pm0.31$ respectively
\citep{Kaneko2006, Sakamoto2009}. All estimated spectral
parameters are given in square brackets in Table~\ref{tab:para}.

Using the spectral parameters and redshift information in
Table~\ref{tab:para} we have calculated the peak isotropic
luminosities for all the bursts in our sample: these results are
shown in Table~\ref{tab:liso}. GRB 080430 has the lowest
luminosity in the sample ($\sim 1.03 \times 10^{ 51}\,\rm
erg\,s^{-1}$), and GRB 080607 has the highest luminosity ($\sim
7.19 \times 10^{ 53}\,\rm erg\,s^{-1}$). The sample spans roughly
three orders of magnitude in luminosity.

We extracted the spectral lags for all combinations of the
canonical BAT energy bands: channel 1 (15--25 keV), 2 (25--50
keV), 3 (50--100 keV) and 4 (100--200 keV). We took the
upper-boundary of channel 4 to be 200 keV because we found that
after the mask weighting the contribution to the light curve from
energies greater than $\sim$200 keV is negligible. The
nomenclature is straightforward, i.e. the spectral lag between
energy channels 4 and 1 is represented by Lag 41. As such there
are six channel combinations and the results for all six are shown
in Table~\ref{tab:lag}. The segment of the light curve used for
the lag extraction ($T+X_{\rm S}$ and $T+X_{\rm E}$, $T$ is the
trigger time), the time binning of the light curve, and the
Gaussian curve fitting range of the CCF vs time delay plot (with
start time, and end time denoted as $LS$ and $LE$ respectively)
are also given in Table~\ref{tab:lag}.

We noticed, as did \cite{WuFenimore2000}, that the lag extraction
is sensitive to a number of parameters. Hence, in
Table~\ref{tab:lag}, we specify the band pass that we used to
extract the lag, segment of the light curve used, temporal bin
resolution, and the fitting range used in the CCF vs time delay
plot. These additional parameters are reported in order to
facilitate reproduction of the results and direct comparison with
other extraction techniques.

\begin{figure}
\epsscale{1.0} \plotone{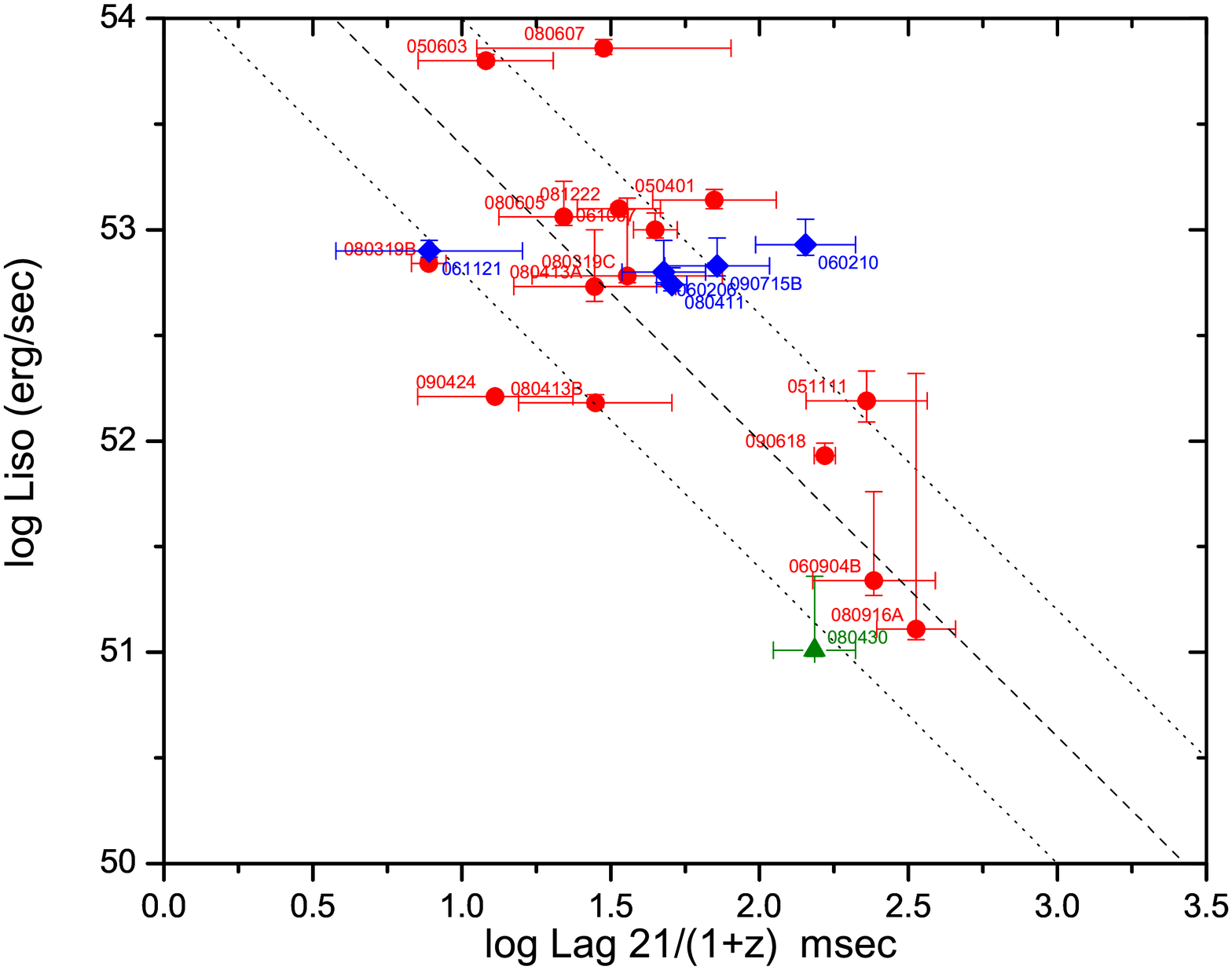} \caption{Isotropic
luminosity as a function of spectral lag between BAT channel 2
(25--50 keV) and 1 (15--25 keV). The ``Gold'', ``Silver'', and
``Bronze'' samples are represented with red circles, blue
diamonds, and green triangles respectively. }\label{Lag21vsLiso}
\end{figure}
\begin{figure}
\epsscale{1.0} \plotone{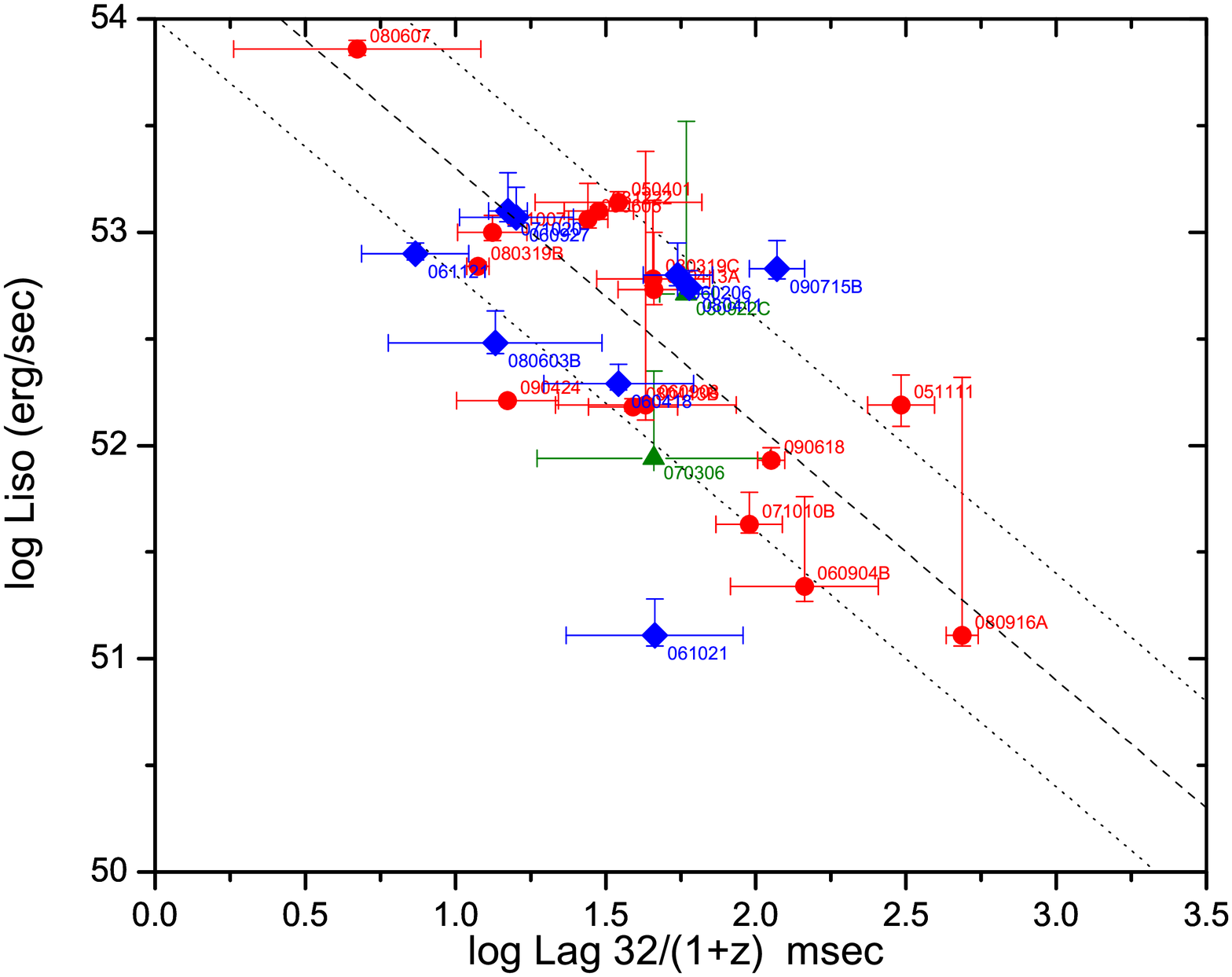} \caption{Isotropic
luminosity as a function of spectral lag between BAT channel 3
(50--100 keV) and 2 (25--50 keV). The ``Gold'', ``Silver'', and
``Bronze'' samples are represented with red circles, blue
diamonds, and green triangles respectively.}\label{Lag32vsLiso}
\end{figure}
\begin{figure}
\epsscale{1.0} \plotone{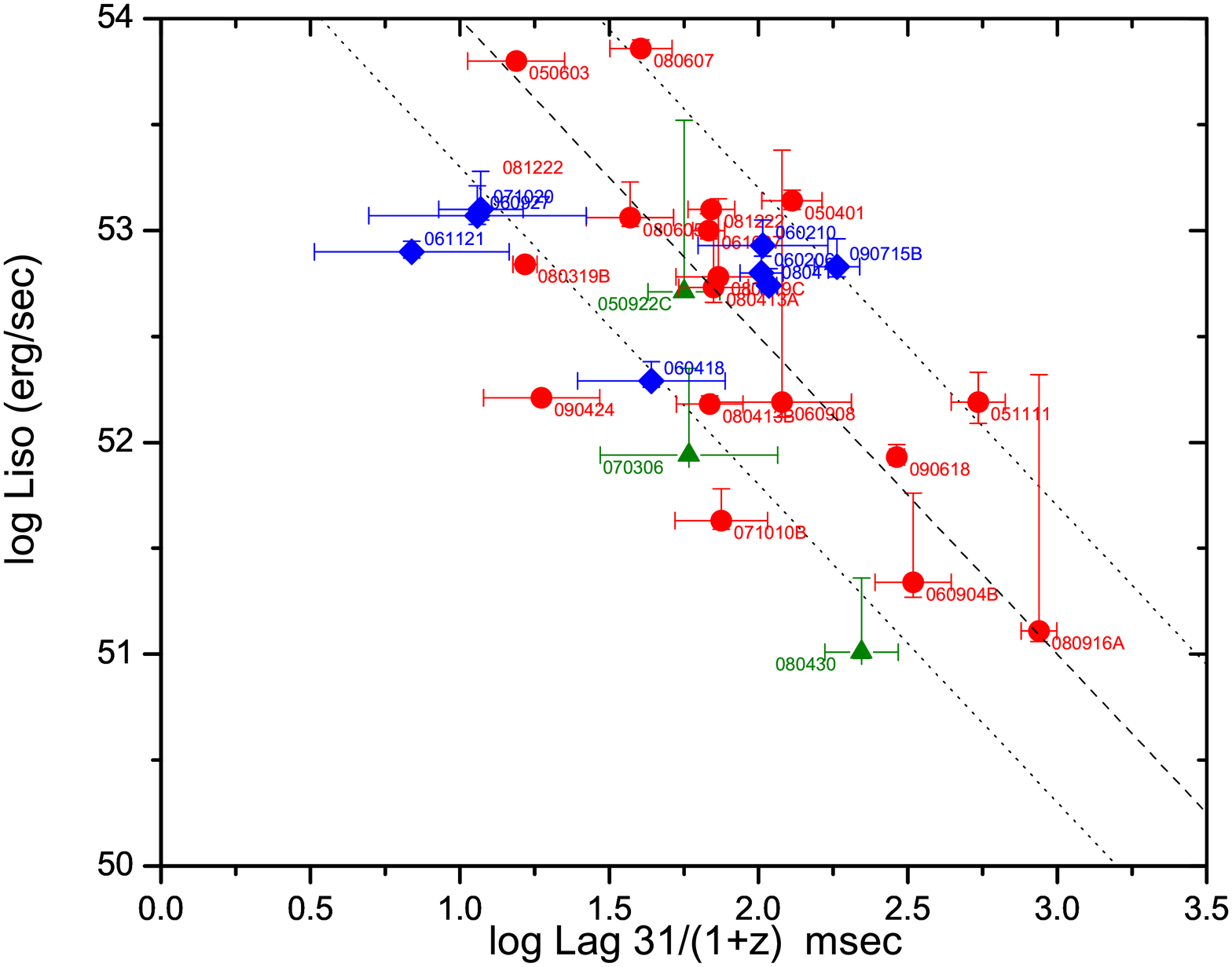} \caption{Isotropic
luminosity as a function of spectral lag between BAT channel 3
(50--100 keV) and 1 (15--25 keV). The ``Gold'', ``Silver'', and
``Bronze'' samples are represented with red circles, blue
diamonds, and green triangles respectively.}\label{Lag31vsLiso}
\end{figure}
\begin{figure}
\epsscale{1.0} \plotone{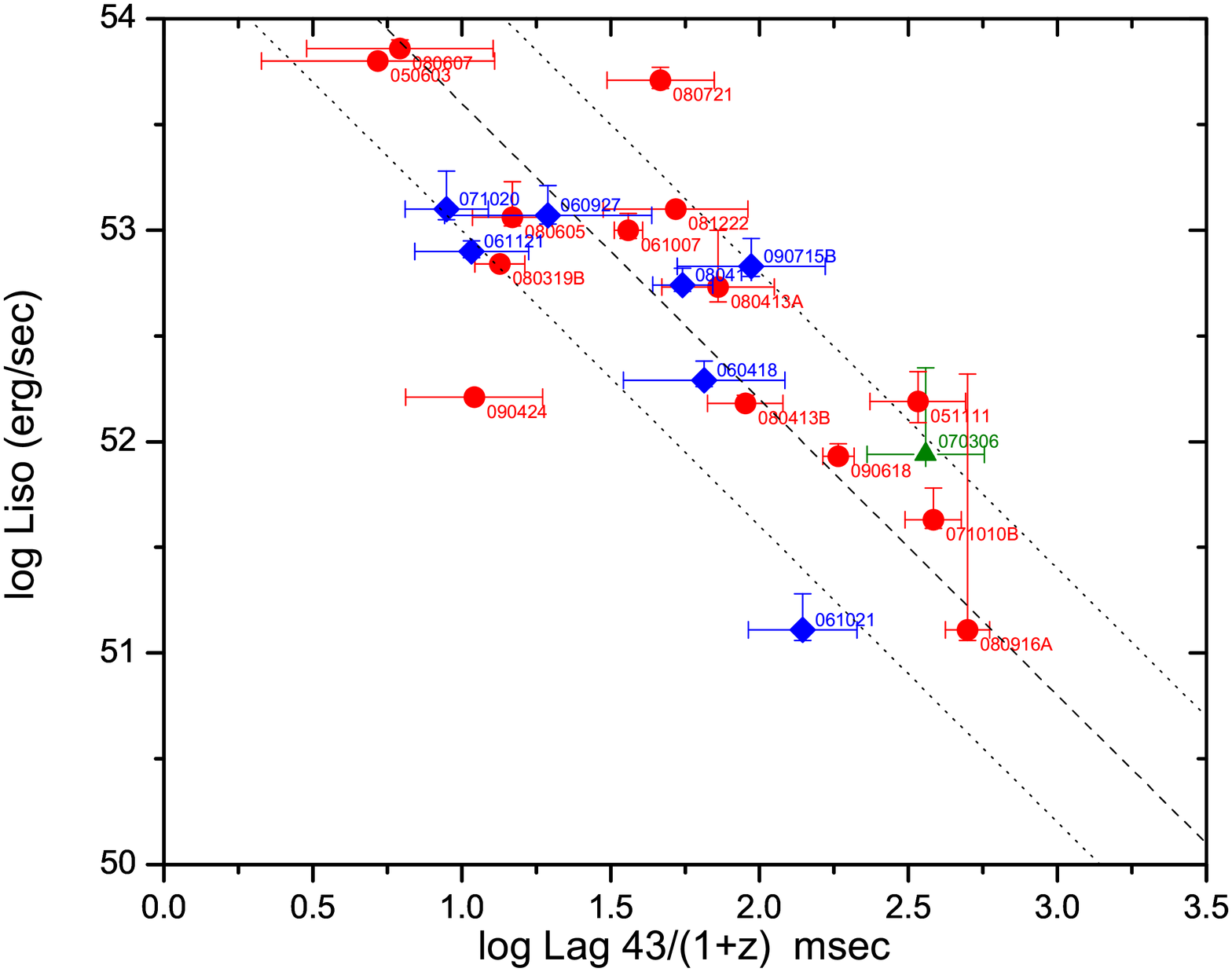} \caption{Isotropic
luminosity as a function of spectral lag between BAT channel 4
(100--200 keV) and 3 (50--100 keV). The ``Gold'', ``Silver'', and
``Bronze'' samples are represented with red circles, blue
diamonds, and green triangles respectively.}\label{Lag43vsLiso}
\end{figure}
\begin{figure}
\epsscale{1.0} \plotone{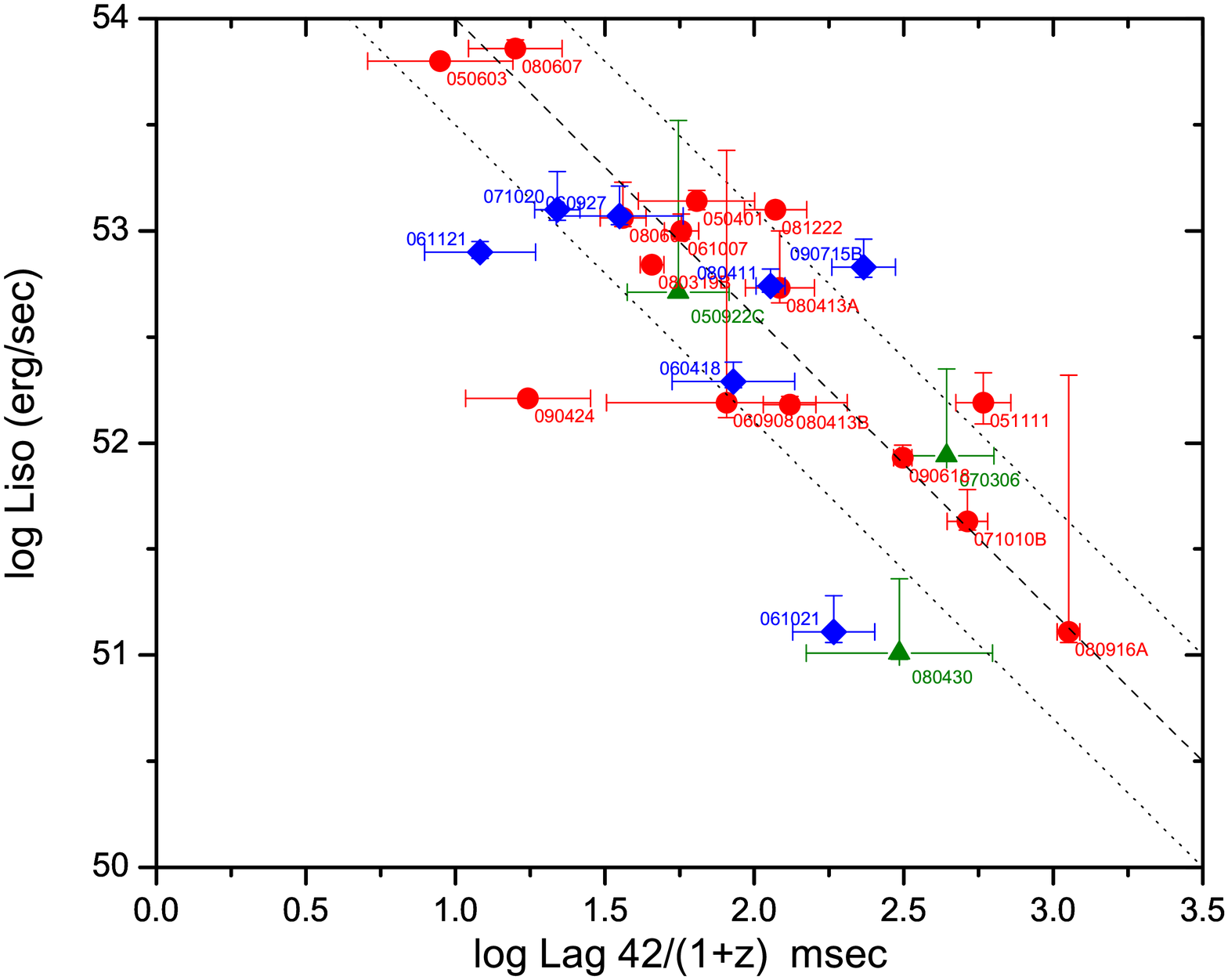} \caption{Isotropic
luminosity as a function of spectral lag between BAT channel 4
(100--200 keV) and 2 (25--50 keV). The ``Gold'', ``Silver'', and
``Bronze'' samples are represented with red circles, blue
diamonds, and green triangles respectively.}\label{Lag42vsLiso}
\end{figure}
\begin{figure}
\epsscale{1.0} \plotone{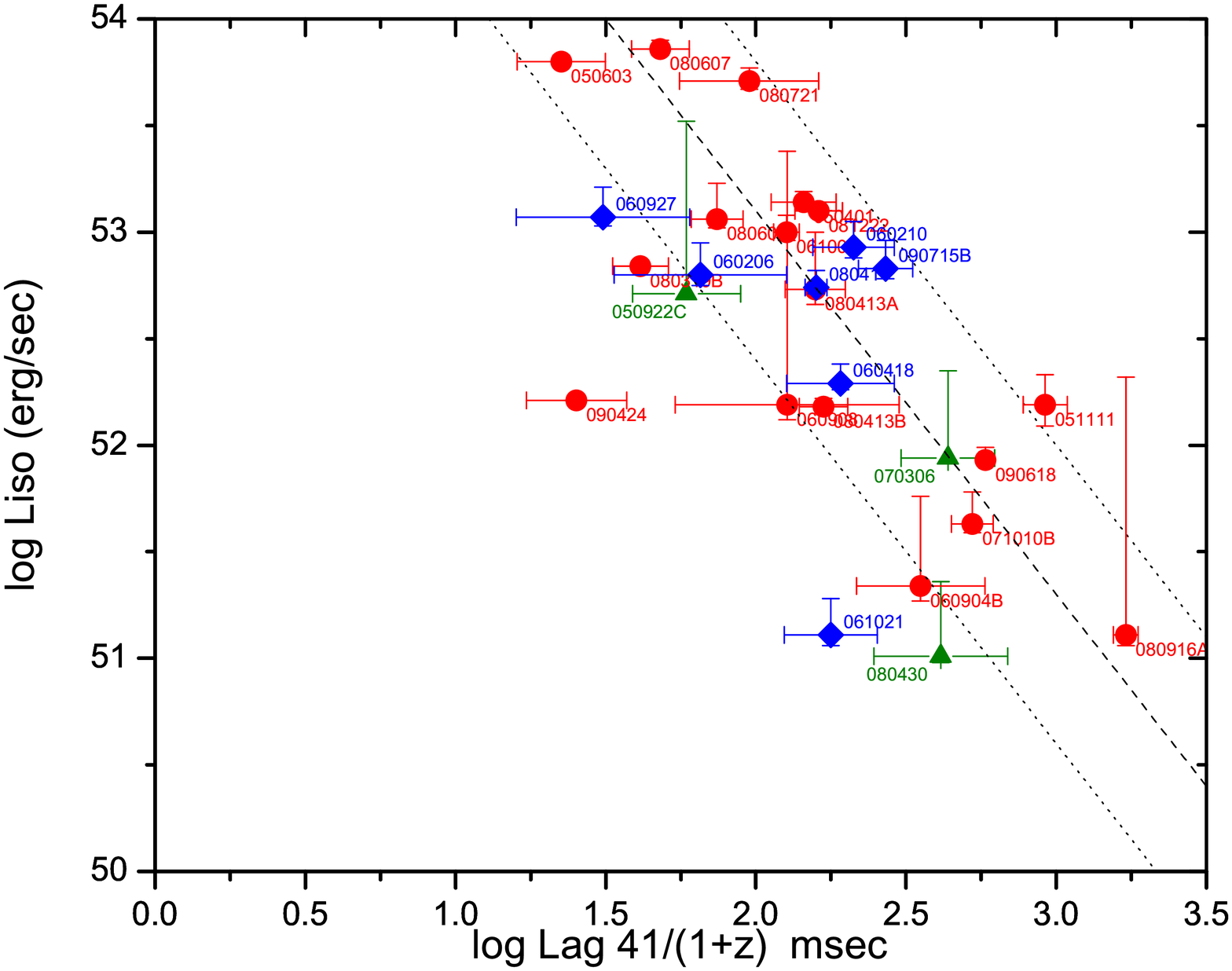} \caption{Isotropic
luminosity as a function of spectral lag between BAT channel 4
(100--200 keV) and 1 (15--25 keV). The ``Gold'', ``Silver'', and
``Bronze'' samples are represented with red circles, blue
diamonds, and green triangles respectively.}\label{Lag41vsLiso}
\end{figure}

Figures~\ref{Lag21vsLiso} through \ref{Lag41vsLiso} show log-log
plots of isotropic peak luminosity vs redshift corrected spectral
lag for various energy channel combinations. Red circles represent
bursts from the ``Gold'' sample, blue diamonds shows bursts from
the ``Silver'' sample and green triangles are bursts from the
``Bronze'' sample. The best-fit power-law curve is also shown in
these plots with a dash line. Since there is a large scatter in
these plots, to compensate, the uncertainties of the fit
parameters are multiplied by a factor of $\sqrt{\rm \chi^2/ndf}$
(see Table~\ref{tab:eband}) \footnote{ndf - number of degrees of
freedom}. The dotted lines indicate the estimated 1 $\sigma$
confidence level, which is obtained from the cumulative fraction
of the residual distribution taken from 16\% to 84\%.

It is interesting to note that GRB 080603B exhibits five negative
lags out of six possible combinations. While these negative lags
are not shown in the plots, it is worth noting that negative lags
are not necessarily unphysical~\citep{Ryde2005}. Moreover, in the
few cases where the uncertainty is large, i.e. the extracted lags
are consistent with zero, these points are not plotted either but
are listed in Table~\ref{tab:lag}. We recognize that the omission
of these negative and zero lags is a potential source of bias.

As seen in Fig.~\ref{Lag21vsLiso} through~\ref{Lag41vsLiso}, our
results support the existence of the lag-luminosity correlation
originally proposed by \cite{norris2000}. Table~\ref{tab:eband}
lists the correlation coefficients for all six channel
combinations. The lag for channel combination 31 has the lowest
correlation with $L_{\rm iso}$, where the correlation coefficient
is -0.60 (with chance probability of $\sim \, 1.5 \times 10^{-3}$)
and the lag for channel 43 has the highest correlation with
coefficient of -0.77 (with chance probability of $\sim \, 3.0
\times 10^{-4}$). However, we note that there is considerable
scatter in the plots. The results of our best fit curves for each
energy band combinations are also given in Table~\ref{tab:eband}.
The mean value of the power-law indices that we get for various
channel combinations is $1.4 \pm 0.3$. Our value is consistent
with the $1.14$ power-law index \cite{norris2000} reported using
lags between BATSE energy bands $100-300$ keV and $25-50$ keV. Our
results are also consistent with \cite{Stamatikos2008} and
\cite{Schaefer2007} who reported values of $1.16 \pm 0.21$ and
$1.01 \pm 0.10$ (assuming an uncertainty of 10\%), respectively.

\section{Discussion} \label{discussion}

\cite{band1997} showed that gamma-ray burst spectra typically
undergo hard-to-soft peak evolution, i.e. the burst peak moves to
later times for lower energy bands. In our sample we have six lag
extractions for each burst. The perfect hard-to-soft peak
evolution scenario is indicated by positive lag values for all
channel combinations plus $\rm lag41 > lag42 > lag43$ and $\rm
lag31 > lag32$. However, all bursts in our sample do not show this
perfect behavior. \cite{band1997} used a scoring method to
quantify the degree of hard-to-soft peak evolution. We used a more
elaborate scoring method to assign a score to each GRB as follows:
First, we increase the burst score by one if one of the six lag
values is positive or decrease it by one if it is negative. Thus,
a GRB can get a score ranging from -6 to +6 at this first step.
Then, we compare the lag values of channel 4 as the base (lag43,
lag42, and lag41). The score is increased by one if the burst
meets one of the following conditions: $\rm lag41
> lag42$, $ \rm lag41 > lag43$ or $\rm lag42 > lag43$. We continue
this procedure for channel 3 as the base also ($\rm lag 31 >
lag32$). We decrease the score by one if it is otherwise.
According to this scoring scheme a score of +10 corresponds to the
perfect case that we mentioned earlier. A positive score indicates
overall hard-to-soft peak evolution in the burst to some degree. A
negative value indicates soft-to-hard peak evolution. Out of 31
bursts in our sample 19 bursts show perfect hard-to-soft peak
evolution with a score of +10. About 97\% of bursts in our sample
have a score of greater than zero, which is consistent with the
90\% value reported by \cite{band1997}.

If one wants to use the lag-luminosity relation as a probe into
the physics of GRBs (in the source rest frame), then a few
corrections to the spectral lag are required; 1) correct for the
time dilation effect (z-correction), 2) take into account the fact
that for GRBs with various redshifts, observed energy bands
correspond to different energy bands at the GRB rest frame
(k-correction). \cite{gehrels2006} approximately corrected
observed spectral lag for the above mentioned effects. We also
examined these corrections. The z-correction is done by
multiplying the lag value by $(1+z)^{-1}$. The k-correction is
approximately done by multiplying the lag value by $(1+z)^{0.33}$
\citep{gehrels2006,Zhang2009}. In table~\ref{tab:corr_correction}
we list the correlation coefficients with no correction, only
z-correction, only k-correction and both corrections applied. For
example correlation coefficient of Lag31 and $L_{\rm iso}$ is
-0.38 without any corrections. After the k-correction the
correlation coefficient is -0.29. Therefore, we do not gain a
significant improvement in the correlation by applying the
k-correction. However, the correlation improves significantly
after the z-correction (-0.60). The approximate k-correction of
\citep{gehrels2006} is based on the assumption that the spectral
lag is proportional to the pulse width and pulse width is
proportional to the energy ~\citep{Zhang2009, Fenimore1995}. These
approximations depend on clearly identifying a pulse in the light
curve and may be of limited validity for multi-pulse structures. A
better method would be to define two energy bands in the GRB rest
frame and project those two bands into the observer frame and
extract lags between them (Ukwatta, et al., 2010 in preparation).

\section{Conclusion} \label{conclusion}

In this work we have used the CCF technique to extract spectral
lags for a sample of $Swift$ BAT GRBs with known redshifts. By
using Monte Carlo simulations, we have extended this technique to
reliably determine the uncertainties in the extracted spectral
lags. Normally these uncertainties would be very difficult to
calculate analytically.

This study provides further support for the existence of the
lag-luminosity correlation, originally proposed by
\cite{norris2000}. We note however, that there is a significant
scatter in the correlation.

The authors are indebted to the late Dr. David L. Band for
fruitful and insightful discussions on the CCF methodology. In
addition, we take this opportunity to acknowledge useful input
from David A. Kahn regarding the luminosity calculations. We also
thank the anonymous referee for comments and suggestions that
significantly improved the paper. The NASA grant NNX08AR44A
provided partial support for this work and is gratefully
acknowledged.

\begin{deluxetable}{lcccccl}
\tabletypesize{\scriptsize} \tablecaption{GRB redshift and
spectral information \label{tab:para}} \tablewidth{0pt}
\tablehead{\colhead{GRB} & \colhead{z} & \colhead{Peak Flux $^{\rm
a}$} & \colhead{$E_{\rm p}$$^{\,b}$} & \colhead{$\alpha^{\,c}$} &
\colhead{$\beta^{\,d}$} & \colhead{Reference} }

\startdata

GRB050401  & 2.899$^{ 1}$  & $10.70\pm0.58$ & $ 119^{+ 16}_{- 16} $   & $0.83^{+0.13}_{-0.13} $ & $2.37^{+0.09}_{-0.09} $ & \cite{2005GCN..3179....1G,Taka2008} \\
GRB050603  & 2.821$^{ 2}$  & $21.50\pm0.67$ & $ 349^{+ 18}_{- 18} $   & $0.79^{+0.04}_{-0.04} $ & $2.15^{+0.06}_{-0.06} $ & \cite{2005GCN..3518....1G,Taka2008} \\
GRB050922C & 2.199$^{ 3}$  & $7.26\pm0.20$  & [$133^{+ 468}_{- 39} $] &[$0.87^{+0.33}_{-0.33}]$ &[$2.36^{+0.31}_{-0.31} $]& \cite{Taka2008} \\
GRB051111  & 1.550$^{ 4}$  & $2.66\pm0.13$  & $ 447^{+ 206}_{- 175} $ & $1.22^{+0.09}_{-0.09} $ & $2.10^{+0.27}_{-4.94} $ & \cite{Hans2009,Taka2008} \\
GRB060206  & 4.056$^{ 5}$  & $2.79\pm0.11$  & $ 75^{+ 12}_{- 12} $    & $1.06^{+0.21}_{-0.21} $ &[$2.36^{+0.31}_{-0.31} $]& \cite{2006GCN..4697....1P,Taka2008} \\
GRB060210  & 3.913$^{ 6}$  & $2.72\pm0.18$  & $ 207^{+ 66}_{- 47} $   & $1.18^{+0.11}_{-0.11} $ &[$2.36^{+0.31}_{-0.31} $]& \cite{Hans2009,Taka2008} \\
GRB060418  & 1.490$^{ 7}$  & $6.52\pm0.22$  & $ 230^{+ 23}_{- 23} $   & $1.50^{+0.09}_{-0.09} $ &[$2.36^{+0.31}_{-0.31} $]& \cite{2006GCN..4989....1G,Taka2008} \\
GRB060904B & 0.703$^{ 8}$  & $2.44\pm0.13$  & $ 103^{+ 59}_{- 26} $   & $0.61^{+0.42}_{-0.42} $ & $1.78^{+0.16}_{-0.23} $ & \cite{Hans2009,Taka2008} \\
GRB060908  & 1.884$^{ 9}$ & $3.03\pm0.16$  & $ 124^{+ 48}_{- 24} $   & $0.89^{+0.20}_{-0.20} $ & $2.24^{+0.34}_{-4.85} $ & \cite{Hans2009,Taka2008} \\
GRB060927  & 5.464$^{ 10}$ & $2.70\pm0.11$  & $ 72^{+ 16}_{- 7} $     & $0.90^{+0.25}_{-0.25} $ &[$2.36^{+0.31}_{-0.31} $]& \cite{Taka2008} \\
GRB061007  & 1.262$^{ 11}$ & $14.60\pm0.23$ & $ 498^{+ 34}_{- 30} $   & $0.53^{+0.06}_{-0.05} $ & $2.61^{+0.16}_{-0.31} $ & \cite{2006GCN..5722....1G,Taka2008} \\
GRB061021  & 0.346$^{ 12}$ & $6.11\pm0.17$  & $ 777^{+ 343}_{- 148} $ & $1.22^{+0.08}_{-0.09} $ &[$2.36^{+0.31}_{-0.31} $]& \cite{2006GCN..5748....1G,Taka2008} \\
GRB061121  & 1.315$^{ 13}$ & $21.10\pm0.29$ & $ 606^{+ 56}_{- 45} $   & $1.32^{+0.02}_{-0.03} $ &[$2.36^{+0.31}_{-0.31} $]& \cite{2006GCN..5837....1G,Taka2008} \\
GRB070306  & 1.496$^{ 14}$ & $4.07\pm0.13$  & [$ 76^{+ 131}_{- 52} $] &[$0.87^{+0.33}_{-0.33}]$ &[$2.36^{+0.31}_{-0.31} $]& \cite{Taka2008} \\
GRB071010B & 0.947$^{ 15}$ & $7.70\pm0.19$  & $ 52^{+ 6}_{- 9} $      & $1.25^{+0.46}_{-0.31} $ & $2.65^{+0.18}_{-0.31} $ & \cite{2007GCN..6879....1G,2007GCNR...92.1} \\
GRB071020  & 2.145$^{ 16}$ & $8.40\pm0.19$  & $ 322^{+ 50}_{- 33} $   & $0.65^{+0.17}_{-0.20} $ &[$2.36^{+0.31}_{-0.31} $]& \cite{2007GCN..6960....1G,2007GCNR...94.2H} \\
GRB080319B & 0.937$^{ 17}$ & $24.80\pm0.31$ & $ 651^{+ 8}_{- 9} $     & $0.82^{+0.01}_{-0.01} $ & $3.87^{+0.28}_{-0.68} $ & \cite{2008GCN..7482....1G,2008GCNR..134.1R} \\
GRB080319C & 1.949$^{ 18}$ & $5.20\pm0.19$  & $ 307^{+ 88}_{- 58} $   & $1.01^{+0.08}_{-0.08} $ & $1.87^{+0.09}_{-0.39} $ & \cite{2008GCN..7487....1G,2008GCN..7483....1S} \\
GRB080411  & 1.030$^{ 19}$ & $43.20\pm0.56$ & $ 259^{+ 22}_{- 17} $   & $1.51^{+0.02}_{-0.03} $ &[$2.36^{+0.31}_{-0.31} $]& \cite{2008GCN..7589....1G,2008GCN..7591....1S} \\
GRB080413A & 2.433$^{ 20}$ & $5.60\pm0.13$  & $ 126^{+ 82}_{- 26} $   & $1.15^{+0.18}_{-0.18} $ & $2.12^{+0.21}_{-4.93} $ & \cite{Hans2009,2008GCNR..129.1M} \\
GRB080413B & 1.101$^{ 21}$ & $18.70\pm0.04$ & $ 67^{+ 8}_{- 5} $      & $1.24^{+0.16}_{-0.16} $ & $2.77^{+0.14}_{-0.17} $ & \cite{Hans2009,2008GCN..7606....1B} \\
GRB080430  & 0.767$^{ 22}$ & $2.60\pm0.13$  & [$ 67^{+ 85}_{- 51} $]  &[$0.87^{+0.33}_{-0.33}]$ &[$2.36^{+0.31}_{-0.31} $]& \cite{2008GCNR..139.1G} \\
GRB080603B & 2.689$^{ 23}$ & $3.50\pm0.13$  & $ 71^{+ 10}_{- 10} $    & $1.21^{+0.19}_{-0.19} $ &[$2.36^{+0.31}_{-0.31} $]& \cite{2008GCNR..144.1M} \\
GRB080605  & 1.640$^{ 24}$ & $19.90\pm0.38$ & $ 297^{+ 29}_{- 25} $   & $0.87^{+0.08}_{-0.08} $ & $2.58^{+0.19}_{-0.53} $ & \cite{2008GCN..7854....1G,2008GCNR..142.1S} \\
GRB080607  & 3.036$^{ 25}$ & $23.10\pm0.69$ & $ 348^{+ 17}_{- 17} $   & $0.76^{+0.04}_{-0.04} $ & $2.57^{+0.11}_{-0.16} $ & \cite{2008GCN..7862....1G,2008GCNR..147.1M} \\
GRB080721  & 2.591$^{ 26}$ & $20.90\pm1.13$ & $ 485^{+ 42}_{- 37} $   & $0.93^{+0.07}_{-0.05} $ & $2.43^{+0.15}_{-0.26} $ & \cite{2008GCN..7995....1G,2008GCNR..156.1M} \\
GRB080916A & 0.689$^{ 27}$ & $2.70\pm0.13$  & $ 121^{+ 50}_{- 16} $   & $0.95^{+0.16}_{-0.16} $ & $2.15^{+0.17}_{-4.91} $ & \cite{Hans2009,2008GCNR..167.3Z} \\
GRB081222  & 2.770$^{ 28}$ & $7.70\pm0.13$  & $ 134^{+ 6}_{- 6} $     & $0.55^{+0.04}_{-0.04} $ & $2.10^{+0.04}_{-0.04} $ & \cite{2008GCN..8715....1B,2009GCNR..190.1G} \\
GRB090424  & 0.544$^{ 29}$ & $71.00\pm1.25$ & $ 177^{+ 2}_{- 2} $     & $0.90^{+0.01}_{-0.01} $ & $2.90^{+0.06}_{-0.06} $ & \cite{2009GCN..9230....1C,2009GCNR..221.1C} \\
GRB090618  & 0.540$^{ 30}$ & $38.80\pm0.50$ & $ 156^{+ 7}_{- 7} $     & $1.26^{+0.04}_{-0.01} $ & $2.50^{+0.09}_{-0.21} $ & \cite{2009GCN..9535....1S,2009GCNR..232.1S} \\
GRB090715B & 3.000$^{ 31}$ & $3.80\pm0.13$  & $ 178^{+ 21}_{- 14} $   & $0.86^{+0.14}_{-0.13} $ &[$2.36^{+0.31}_{-0.31} $]& \cite{2009GCN..9679....1S,2009GCNR..236.1V} \\

\enddata \tablerefs{
(1) \cite{Watson2006}; (2) \cite{2005GCN..3520....1B}; (3)
\cite{Piranomonte2008}; (4) \cite{Penprase2006}; (5)
\cite{Fynbo2009}; (6) \cite{Fynbo2009}; (7)
\cite{2006GCN..5002....1P}; (8) \cite{Fynbo2009}; (9)
\cite{Fynbo2009}; (10) \cite{Fynbo2009}; (11) \cite{Fynbo2009};
(12) \cite{Fynbo2009}; (13) \cite{Fynbo2009}; (14)
\cite{Jaunsen2008}; (15) \cite{Cenko2007}; (16)
\cite{2007GCN..6952....1J}; (17) \cite{DElia2009}; (18)
\cite{Fynbo2009}; (19) \cite{Fynbo2009}; (20) \cite{Fynbo2009};
(21) \cite{Fynbo2009}; (22) \cite{2008GCN..7654....1C}; (23)
\cite{Fynbo2009}; (24) \cite{Fynbo2009}; (25)
\cite{Prochaska2009}; (26) \cite{Fynbo2009}; (27)
\cite{Fynbo2009}; (28) \cite{Cucchiara2008}; (29)
\cite{2009GCN..9243....1C}; (30) \cite{2009GCN..9518....1S}; (31)
\cite{2009GCN..9673....1S}. }

\tablenotetext{a}{1-second peak photon flux measured in $\rm
photon\, \,cm^{-2} \, s^{-1}$ in the energy range $15-150$ keV.}

\tablenotetext{b}{Peak energy is given in keV. Values in brackets
indicates estimated values using the method described in
\cite{Sakamoto2009}.}

\tablenotetext{c}{Values in brackets indicates estimated
high-energy photon index, $\alpha$, which is the mean value of the
BATSE $\alpha$ distribution \citep{Kaneko2006,Sakamoto2009}.}

\tablenotetext{d}{Values in brackets indicates estimated
high-energy photon index, $\beta$, which is the mean value of the
BATSE $\beta$ distribution \citep{Kaneko2006,Sakamoto2009}.}

\tablecomments{Note that uncertainties of parameters that are
reported with 90\% confidence level have been reduced to $1
\sigma$ level for consistency.}
\end{deluxetable}

\begin{deluxetable}{lcrlcr}
\tabletypesize{\scriptsize} \tablecaption{GRB redshift and
calculated isotropic peak luminosity \label{tab:liso}}
\tablewidth{0pt} \tablehead{\colhead{GRB} & \colhead{Redshift} &
\colhead{Peak Isotropic Luminosity $^{\rm a}$} & \colhead{GRB} &
\colhead{Redshift} & \colhead{Peak Isotropic Luminosity $^{\rm
a}$} }

\startdata

GRB050401 & 2.899 & ($1.38^{+0.16}_{-0.13} ) \times 10^{ 53}$  & GRB080319B & 0.937 & ($6.96^{+0.32}_{-0.14} ) \times 10^{ 52}$ \\
GRB050603 & 2.821 & ($6.32^{+0.47}_{-0.34} ) \times 10^{ 53}$  & GRB080319C & 1.949 & ($6.04^{+8.04}_{-0.42} ) \times 10^{ 52}$ \\
GRB050922C & 2.199 & ($5.17^{+28.00}_{-0.01} ) \times 10^{ 52}$  & GRB080411 & 1.030 & ($5.49^{+1.11}_{-0.34} ) \times 10^{ 52}$ \\
GRB051111 & 1.550 & ($1.55^{+0.61}_{-0.33} ) \times 10^{ 52}$  & GRB080413A & 2.433 & ($5.38^{+4.69}_{-0.83} ) \times 10^{ 52}$ \\
GRB060206 & 4.056 & ($6.28^{+2.50}_{-0.62} ) \times 10^{ 52}$  & GRB080413B & 1.101 & ($1.51^{+0.15}_{-0.06} ) \times 10^{ 52}$ \\
GRB060210 & 3.913 & ($8.53^{+2.75}_{-0.92} ) \times 10^{ 52}$  & GRB080430 & 0.767 & ($1.03^{+1.30}_{-0.07} ) \times 10^{ 51}$ \\
GRB060418 & 1.490 & ($1.96^{+0.43}_{-0.13} ) \times 10^{ 52}$  & GRB080603B & 2.689 & ($2.99^{+1.25}_{-0.30} ) \times 10^{ 52}$ \\
GRB060904B & 0.703 & ($2.18^{+3.59}_{-0.32} ) \times 10^{ 51}$  & GRB080605 & 1.640 & ($1.15^{+0.56}_{-0.09} ) \times 10^{ 53}$ \\
GRB060908 & 1.884 & ($1.54^{+22.50}_{-0.22} ) \times 10^{ 52}$  & GRB080607 & 3.036 & ($7.19^{+0.64}_{-0.41} ) \times 10^{ 53}$ \\
GRB060927 & 5.464 & ($1.17^{+0.43}_{-0.10} ) \times 10^{ 53}$  & GRB080721 & 2.591 & ($5.18^{+0.83}_{-0.47} ) \times 10^{ 53}$ \\
GRB061007 & 1.262 & ($1.01^{+0.20}_{-0.08} ) \times 10^{ 53}$  & GRB080916A & 0.689 & ($1.30^{+19.90}_{-0.15} ) \times 10^{ 51}$ \\
GRB061021 & 0.346 & ($1.30^{+0.60}_{-0.13} ) \times 10^{ 51}$  & GRB081222 & 2.770 & ($1.26^{+0.07}_{-0.06} ) \times 10^{ 53}$ \\
GRB061121 & 1.315 & ($7.89^{+1.02}_{-0.47} ) \times 10^{ 52}$  & GRB090424 & 0.544 & ($1.62^{+0.05}_{-0.04} ) \times 10^{ 52}$ \\
GRB070306 & 1.496 & ($8.67^{+13.50}_{-0.27} ) \times 10^{ 51}$  & GRB090618 & 0.540 & ($8.47^{+1.17}_{-0.34} ) \times 10^{ 51}$ \\
GRB071010B & 0.947 & ($4.24^{+1.72}_{-0.33} ) \times 10^{ 51}$  & GRB090715B & 3.000 & ($6.79^{+2.42}_{-0.71} ) \times 10^{ 52}$ \\
GRB071020 & 2.145 & ($1.27^{+0.64}_{-0.15} ) \times 10^{ 53}$  &  $$ \\

\enddata
\tablenotetext{a}{Isotropic equivalent peak photon luminosity in
$\rm erg \, s^{-1}$ between GRB rest frame energy range 1 and
10,000 keV as described in Section~\ref{methodology}.}

\end{deluxetable}

\begin{deluxetable}{lccrrrccc}
\tabletypesize{\scriptsize} \tablecaption{Spectral lag values of
long duration $Swift$ BAT GRBs \label{tab:lag}} \tablewidth{0pt}
\tablehead{\colhead{GRB} & \colhead{Trigger ID} & \colhead{LagXX}
& \colhead{$T+X_{\rm S}$ (s)} & \colhead{$T+X_{\rm E}$ (s)} &
\colhead{Bin Size (ms)} & \colhead{LS (s)} & \colhead{LE (s)}&
\colhead{Lag Value (ms)}} \startdata

GRB050401 & 113120 & Lag 21 & 23.03 & 29.43 & 32 & -1.00 & 1.50 &  $275 \pm 131$ \\
 &  & Lag 31 & 23.03 & 29.43 & 32 & -1.00 & 2.00 &  $504 \pm 117$ \\
 &  & Lag 41 & 23.03 & 29.43 & 64 & -1.00 & 2.00 &  $562 \pm 140$ \\
 &  & Lag 32 & 23.03 & 29.43 & 16 & -1.00 & 1.00 &  $136 \pm 87$ \\
 &  & Lag 42 & 23.03 & 29.43 & 64 & -1.50 & 1.50 &  $250 \pm 112$ \\
 &  & Lag 43 & 23.03 & 29.43 & 64 & -2.00 & 2.00 &  $106 \pm 118$ \\
GRB050603 & 131560 & Lag 21 & -3.83 & 3.08 & 8 & -0.40 & 0.40 &  $46 \pm 24$ \\
 &  & Lag 31 & -3.83 & 3.08 & 8 & -0.40 & 0.40 &  $59 \pm 22$ \\
 &  & Lag 41 & -3.83 & 3.08 & 16 & -0.40 & 0.40 &  $86 \pm 29$ \\
 &  & Lag 32 & -3.83 & 3.08 & 4 & -0.20 & 0.20 &  $4 \pm 11$ \\
 &  & Lag 42 & -3.83 & 3.08 & 16 & -0.40 & 0.40 &  $34 \pm 19$ \\
 &  & Lag 43 & -3.83 & 3.08 & 16 & -0.50 & 0.50 &  $20 \pm 18$ \\
GRB050922C & 156467 & Lag 21 & -2.70 & 2.94 & 8 & -0.40 & 0.40 &  $9 \pm 35$ \\
 &  & Lag 31 & -2.70 & 2.94 & 8 & -1.00 & 1.00 &  $180 \pm 50$ \\
 &  & Lag 41 & -2.70 & 2.94 & 16 & -1.00 & 1.00 &  $188 \pm 78$ \\
 &  & Lag 32 & -2.70 & 2.94 & 4 & -1.00 & 1.00 &  $188 \pm 39$ \\
 &  & Lag 42 & -2.70 & 2.94 & 16 & -1.00 & 1.00 &  $178 \pm 70$ \\
 &  & Lag 43 & -2.70 & 2.94 & 16 & -1.00 & 1.00 &  $19 \pm 72$ \\
GRB051111 & 163438 & Lag 21 & -6.96 & 28.62 & 32 & -5.00 & 4.00 &  $583 \pm 273$ \\
 &  & Lag 31 & -6.96 & 28.62 & 32 & -4.00 & 4.00 &  $1383 \pm 288$ \\
 &  & Lag 41 & -6.96 & 28.62 & 128 & -4.00 & 8.00 &  $2343 \pm 397$ \\
 &  & Lag 32 & -6.96 & 28.62 & 16 & -5.00 & 4.00 &  $776 \pm 200$ \\
 &  & Lag 42 & -6.96 & 28.62 & 64 & -5.00 & 5.00 &  $1486 \pm 314$ \\
 &  & Lag 43 & -6.96 & 28.62 & 64 & -5.00 & 5.00 &  $866 \pm 319$ \\
GRB060206 & 180455 & Lag 21 & -1.29 & 8.18 & 8 & -1.50 & 1.50 &  $241 \pm 78$ \\
 &  & Lag 31 & -1.29 & 8.18 & 16 & -1.00 & 2.00 &  $517 \pm 85$ \\
 &  & Lag 41 & -1.29 & 8.18 & 64 & -1.50 & 2.00 &  $331 \pm 219$ \\
 &  & Lag 32 & -1.29 & 8.18 & 8 & -1.50 & 1.50 &  $278 \pm 74$ \\
 &  & Lag 42 & -1.29 & 8.18 & 64 & -1.50 & 1.50 &  $82 \pm 193$ \\
 &  & Lag 43 & -1.29 & 8.18 & 64 & -2.00 & 2.00 &  $-163 \pm 189$ \\
GRB060210 & 180977 & Lag 21 & -3.37 & 5.08 & 64 & -5.00 & 4.00 &  $700 \pm 270$ \\
 &  & Lag 31 & -3.37 & 5.08 & 64 & -5.00 & 4.00 &  $508 \pm 254$ \\
 &  & Lag 41 & -3.37 & 5.08 & 256 & -5.00 & 4.00 &  $1038 \pm 324$ \\
 &  & Lag 32 & -3.37 & 5.08 & 64 & -4.00 & 4.00 &  $-175 \pm 174$ \\
 &  & Lag 42 & -3.37 & 5.08 & 128 & -4.00 & 4.00 &  $98 \pm 225$ \\
 &  & Lag 43 & -3.37 & 5.08 & 256 & -5.00 & 2.00 &  $34 \pm 195$ \\
GRB060418 & 205851 & Lag 21 & -7.66 & 33.04 & 16 & -2.00 & 2.00 &  $22 \pm 62$ \\
 &  & Lag 31 & -7.66 & 33.04 & 32 & -2.00 & 2.00 &  $109 \pm 62$ \\
 &  & Lag 41 & -7.66 & 33.04 & 128 & -2.00 & 2.00 &  $476 \pm 196$ \\
 &  & Lag 32 & -7.66 & 33.04 & 16 & -2.00 & 2.00 &  $87 \pm 50$ \\
 &  & Lag 42 & -7.66 & 33.04 & 64 & -1.00 & 1.00 &  $212 \pm 100$ \\
 &  & Lag 43 & -7.66 & 33.04 & 64 & -1.00 & 1.00 &  $162 \pm 101$ \\
GRB060904B & 228006 & Lag 21 & -1.97 & 10.32 & 32 & -2.00 & 2.00 &  $412 \pm 195$ \\
 &  & Lag 31 & -1.97 & 10.32 & 32 & -2.00 & 2.00 &  $560 \pm 164$ \\
 &  & Lag 41 & -1.97 & 10.32 & 128 & -3.00 & 3.00 &  $602 \pm 296$ \\
 &  & Lag 32 & -1.97 & 10.32 & 32 & -2.00 & 2.00 &  $247 \pm 140$ \\
 &  & Lag 42 & -1.97 & 10.32 & 128 & -2.50 & 3.00 &  $175 \pm 292$ \\
 &  & Lag 43 & -1.97 & 10.32 & 128 & -3.00 & 3.00 &  $32 \pm 273$ \\
GRB060908 & 228581 & Lag 21 & -10.91 & 3.68 & 32 & -2.00 & 2.00 &  $118 \pm 142$ \\
 &  & Lag 31 & -10.91 & 3.68 & 32 & -2.00 & 2.00 &  $346 \pm 185$ \\
 &  & Lag 41 & -10.91 & 3.68 & 128 & -4.00 & 4.00 &  $367 \pm 315$ \\
 &  & Lag 32 & -10.91 & 3.68 & 16 & -2.00 & 2.00 &  $124 \pm 86$ \\
 &  & Lag 42 & -10.91 & 3.68 & 64 & -2.00 & 2.00 &  $233 \pm 216$ \\
 &  & Lag 43 & -10.91 & 3.68 & 128 & -4.00 & 4.00 &  $134 \pm 253$ \\
GRB060927 & 231362 & Lag 21 & -1.69 & 8.04 & 16 & -0.60 & 0.60 &  $9 \pm 46$ \\
 &  & Lag 31 & -1.69 & 8.04 & 64 & -1.00 & 1.00 &  $74 \pm 62$ \\
 &  & Lag 41 & -1.69 & 8.04 & 256 & -1.50 & 1.50 &  $200 \pm 133$ \\
 &  & Lag 32 & -1.69 & 8.04 & 16 & -1.00 & 1.00 &  $103 \pm 45$ \\
 &  & Lag 42 & -1.69 & 8.04 & 128 & -1.20 & 1.50 &  $229 \pm 112$ \\
 &  & Lag 43 & -1.69 & 8.04 & 128 & -1.20 & 1.50 &  $126 \pm 101$ \\
GRB061007 & 232683 & Lag 21 & 23.86 & 65.08 & 2 & -0.30 & 0.50 &  $101 \pm 17$ \\
 &  & Lag 31 & 23.86 & 65.08 & 2 & -0.30 & 0.50 &  $154 \pm 19$ \\
 &  & Lag 41 & 23.86 & 65.08 & 4 & -0.50 & 0.80 &  $286 \pm 28$ \\
 &  & Lag 32 & 23.86 & 65.08 & 2 & -0.20 & 0.20 &  $30 \pm 8$ \\
 &  & Lag 42 & 23.86 & 65.08 & 2 & -0.40 & 0.40 &  $129 \pm 17$ \\
 &  & Lag 43 & 23.86 & 65.08 & 2 & -0.30 & 0.40 &  $82 \pm 9$ \\
GRB061021 & 234905 & Lag 21 & -0.46 & 14.64 & 8 & -1.00 & 1.00 &  $-25 \pm 52$ \\
 &  & Lag 31 & -0.46 & 14.64 & 8 & -1.00 & 1.00 &  $49 \pm 51$ \\
 &  & Lag 41 & -0.46 & 14.64 & 32 & -1.60 & 1.60 &  $239 \pm 85$ \\
 &  & Lag 32 & -0.46 & 14.64 & 8 & -1.00 & 1.00 &  $62 \pm 42$ \\
 &  & Lag 42 & -0.46 & 14.64 & 32 & -1.00 & 1.20 &  $248 \pm 78$ \\
 &  & Lag 43 & -0.46 & 14.64 & 32 & -1.00 & 1.20 &  $188 \pm 79$ \\
GRB061121 & 239899 & Lag 21 & 60.44 & 80.66 & 1 & -0.20 & 0.20 &  $18 \pm 13$ \\
 &  & Lag 31 & 60.44 & 80.66 & 1 & -0.20 & 0.20 &  $16 \pm 12$ \\
 &  & Lag 41 & 60.44 & 80.66 & 4 & -0.40 & 0.40 &  $26 \pm 26$ \\
 &  & Lag 32 & 60.44 & 80.66 & 1 & -0.20 & 0.25 &  $17 \pm 7$ \\
 &  & Lag 42 & 60.44 & 80.66 & 2 & -0.20 & 0.25 &  $28 \pm 12$ \\
 &  & Lag 43 & 60.44 & 80.66 & 2 & -0.20 & 0.25 &  $25 \pm 11$ \\
GRB070306 & 263361 & Lag 21 & 90.00 & 118.42 & 8 & -2.00 & 2.00 &  $88 \pm 106$ \\
 &  & Lag 31 & 90.00 & 118.42 & 16 & -2.00 & 2.00 &  $146 \pm 100$ \\
 &  & Lag 41 & 90.00 & 118.42 & 64 & -4.00 & 6.00 &  $1088 \pm 391$ \\
 &  & Lag 32 & 90.00 & 118.42 & 8 & -2.00 & 2.00 &  $114 \pm 102$ \\
 &  & Lag 42 & 90.00 & 118.42 & 64 & -4.00 & 6.00 &  $1098 \pm 399$ \\
 &  & Lag 43 & 90.00 & 118.42 & 64 & -4.00 & 6.00 &  $900 \pm 408$ \\
GRB071010B & 293795 & Lag 21 & -1.70 & 17.24 & 2 & -1.00 & 1.00 &  $-26 \pm 48$ \\
 &  & Lag 31 & -1.70 & 17.24 & 4 & -1.00 & 1.00 &  $146 \pm 52$ \\
 &  & Lag 41 & -1.70 & 17.24 & 32 & -2.00 & 4.00 &  $1024 \pm 163$ \\
 &  & Lag 32 & -1.70 & 17.24 & 4 & -1.00 & 1.00 &  $185 \pm 47$ \\
 &  & Lag 42 & -1.70 & 17.24 & 32 & -2.00 & 4.00 &  $1005 \pm 157$ \\
 &  & Lag 43 & -1.70 & 17.24 & 32 & -2.00 & 4.00 &  $745 \pm 161$ \\
GRB071020 & 294835 & Lag 21 & -3.22 & 1.14 & 2 & -0.10 & 0.15 &  $7 \pm 7$ \\
 &  & Lag 31 & -3.22 & 1.14 & 2 & -0.10 & 0.20 &  $37 \pm 12$ \\
 &  & Lag 41 & -3.22 & 1.14 & 8 & -0.50 & 0.50 &  $-50 \pm 30$ \\
 &  & Lag 32 & -3.22 & 1.14 & 2 & -0.10 & 0.25 &  $47 \pm 7$ \\
 &  & Lag 42 & -3.22 & 1.14 & 4 & -0.10 & 0.30 &  $69 \pm 12$ \\
 &  & Lag 43 & -3.22 & 1.14 & 4 & -0.20 & 0.30 &  $28 \pm 9$ \\
GRB080319B & 306757 & Lag 21 & -2.85 & 57.57 & 2 & -0.10 & 0.14 &  $15 \pm 2$ \\
 &  & Lag 31 & -2.85 & 57.57 & 2 & -0.10 & 0.14 &  $32 \pm 3$ \\
 &  & Lag 41 & -2.85 & 57.57 & 2 & -0.20 & 0.20 &  $80 \pm 17$ \\
 &  & Lag 32 & -2.85 & 57.57 & 2 & -0.10 & 0.14 &  $23 \pm 2$ \\
 &  & Lag 42 & -2.85 & 57.57 & 2 & -0.20 & 0.30 &  $88 \pm 8$ \\
 &  & Lag 43 & -2.85 & 57.57 & 2 & -0.20 & 0.20 &  $26 \pm 5$ \\
GRB080319C & 306778 & Lag 21 & -0.77 & 13.31 & 16 & -1.00 & 1.00 &  $106 \pm 78$ \\
 &  & Lag 31 & -0.77 & 13.31 & 16 & -2.00 & 2.00 &  $216 \pm 70$ \\
 &  & Lag 41 & -0.77 & 13.31 & 64 & -2.00 & 2.00 &  $89 \pm 132$ \\
 &  & Lag 32 & -0.77 & 13.31 & 16 & -1.00 & 1.00 &  $134 \pm 58$ \\
 &  & Lag 42 & -0.77 & 13.31 & 32 & -1.00 & 1.00 &  $-77 \pm 95$ \\
 &  & Lag 43 & -0.77 & 13.31 & 32 & -1.00 & 1.00 &  $-119 \pm 99$ \\
GRB080411 & 309010 & Lag 21 & 38.46 & 48.45 & 2 & -1.00 & 1.00 &  $103 \pm 12$ \\
 &  & Lag 31 & 38.46 & 48.45 & 2 & -1.00 & 1.00 &  $220 \pm 13$ \\
 &  & Lag 41 & 38.46 & 48.45 & 2 & -1.00 & 1.00 &  $322 \pm 27$ \\
 &  & Lag 32 & 38.46 & 48.45 & 2 & -1.00 & 1.00 &  $122 \pm 11$ \\
 &  & Lag 42 & 38.46 & 48.45 & 2 & -1.00 & 1.00 &  $230 \pm 26$ \\
 &  & Lag 43 & 38.46 & 48.45 & 2 & -1.00 & 1.00 &  $112 \pm 26$ \\
GRB080413A & 309096 & Lag 21 & -0.42 & 9.05 & 8 & -1.00 & 1.00 &  $96 \pm 60$ \\
 &  & Lag 31 & -0.42 & 9.05 & 8 & -1.00 & 1.00 &  $242 \pm 65$ \\
 &  & Lag 41 & -0.42 & 9.05 & 64 & -1.00 & 2.00 &  $542 \pm 125$ \\
 &  & Lag 32 & -0.42 & 9.05 & 8 & -1.00 & 1.00 &  $157 \pm 43$ \\
 &  & Lag 42 & -0.42 & 9.05 & 32 & -1.00 & 2.00 &  $418 \pm 111$ \\
 &  & Lag 43 & -0.42 & 9.05 & 32 & -1.00 & 2.00 &  $249 \pm 108$ \\
GRB080413B & 309111 & Lag 21 & -1.44 & 4.96 & 8 & -1.00 & 1.00 &  $59 \pm 35$ \\
 &  & Lag 31 & -1.44 & 4.96 & 8 & -1.00 & 1.00 &  $144 \pm 37$ \\
 &  & Lag 41 & -1.44 & 4.96 & 16 & -1.00 & 1.00 &  $353 \pm 66$ \\
 &  & Lag 32 & -1.44 & 4.96 & 8 & -1.00 & 1.00 &  $82 \pm 28$ \\
 &  & Lag 42 & -1.44 & 4.96 & 16 & -1.00 & 1.00 &  $276 \pm 56$ \\
 &  & Lag 43 & -1.44 & 4.96 & 16 & -1.00 & 1.00 &  $188 \pm 55$ \\
GRB080430 & 310613 & Lag 21 & -1.24 & 12.84 & 32 & -2.00 & 2.00 &  $270 \pm 86$ \\
 &  & Lag 31 & -1.24 & 12.84 & 32 & -2.00 & 2.00 &  $391 \pm 109$ \\
 &  & Lag 41 & -1.24 & 12.84 & 256 & -4.00 & 4.00 &  $730 \pm 374$ \\
 &  & Lag 32 & -1.24 & 12.84 & 32 & -2.00 & 2.00 &  $83 \pm 100$ \\
 &  & Lag 42 & -1.24 & 12.84 & 256 & -4.00 & 4.00 &  $540 \pm 387$ \\
 &  & Lag 43 & -1.24 & 12.84 & 256 & -4.00 & 4.00 &  $388 \pm 397$ \\
GRB080603B & 313087 & Lag 21 & -0.54 & 5.10 & 16 & -1.00 & 1.00 &  $-222 \pm 61$ \\
 &  & Lag 31 & -0.54 & 5.10 & 16 & -1.00 & 1.00 &  $-197 \pm 67$ \\
 &  & Lag 41 & -0.54 & 5.10 & 32 & -1.00 & 1.00 &  $-427 \pm 163$ \\
 &  & Lag 32 & -0.54 & 5.10 & 16 & -1.00 & 1.00 &  $50 \pm 41$ \\
 &  & Lag 42 & -0.54 & 5.10 & 32 & -1.00 & 0.50 &  $-103 \pm 71$ \\
 &  & Lag 43 & -0.54 & 5.10 & 32 & -1.00 & 0.50 &  $-172 \pm 56$ \\
GRB080605 & 313299 & Lag 21 & -5.46 & 15.53 & 4 & -1.00 & 1.00 &  $58 \pm 29$ \\
 &  & Lag 31 & -5.46 & 15.53 & 4 & -1.00 & 1.00 &  $98 \pm 33$ \\
 &  & Lag 41 & -5.46 & 15.53 & 16 & -0.50 & 1.20 &  $196 \pm 39$ \\
 &  & Lag 32 & -5.46 & 15.53 & 2 & -0.30 & 0.40 &  $73 \pm 11$ \\
 &  & Lag 42 & -5.46 & 15.53 & 8 & -0.30 & 0.40 &  $96 \pm 17$ \\
 &  & Lag 43 & -5.46 & 15.53 & 8 & -0.30 & 0.40 &  $39 \pm 12$ \\
GRB080607 & 313417 & Lag 21 & -6.13 & 12.05 & 8 & -0.40 & 0.40 &  $121 \pm 119$ \\
 &  & Lag 31 & -6.13 & 12.05 & 8 & -0.40 & 0.60 &  $163 \pm 39$ \\
 &  & Lag 41 & -6.13 & 12.05 & 16 & -0.40 & 0.60 &  $194 \pm 43$ \\
 &  & Lag 32 & -6.13 & 12.05 & 8 & -0.40 & 0.40 &  $19 \pm 17$ \\
 &  & Lag 42 & -6.13 & 12.05 & 8 & -0.40 & 0.40 &  $64 \pm 23$ \\
 &  & Lag 43 & -6.13 & 12.05 & 8 & -0.40 & 0.40 &  $25 \pm 18$ \\
GRB080721 & 317508 & Lag 21 & -3.39 & 8.64 & 64 & -2.00 & 2.00 &  $99 \pm 149$ \\
 &  & Lag 31 & -3.39 & 8.64 & 64 & -2.00 & 2.00 &  $122 \pm 138$ \\
 &  & Lag 41 & -3.39 & 8.64 & 128 & -2.00 & 2.00 &  $341 \pm 182$ \\
 &  & Lag 32 & -3.39 & 8.64 & 16 & -0.80 & 0.80 &  $16 \pm 58$ \\
 &  & Lag 42 & -3.39 & 8.64 & 32 & -0.80 & 0.80 &  $256 \pm 308$ \\
 &  & Lag 43 & -3.39 & 8.64 & 32 & -0.80 & 0.80 &  $167 \pm 69$ \\
GRB080916A & 324895 & Lag 21 & -2.66 & 39.58 & 16 & -2.00 & 3.00 &  $566 \pm 172$ \\
 &  & Lag 31 & -2.66 & 39.58 & 32 & -2.00 & 4.00 &  $1468 \pm 202$ \\
 &  & Lag 41 & -2.66 & 39.58 & 256 & -4.00 & 6.00 &  $2879 \pm 271$ \\
 &  & Lag 32 & -2.66 & 39.58 & 32 & -4.00 & 4.00 &  $821 \pm 100$ \\
 &  & Lag 42 & -2.66 & 39.58 & 128 & -2.00 & 6.00 &  $1900 \pm 165$ \\
 &  & Lag 43 & -2.66 & 39.58 & 64 & -2.00 & 4.00 &  $842 \pm 143$ \\
GRB081222 & 337914 & Lag 21 & -0.80 & 15.58 & 2 & -0.80 & 1.20 &  $127 \pm 41$ \\
 &  & Lag 31 & -0.80 & 15.58 & 2 & -0.80 & 1.00 &  $262 \pm 47$ \\
 &  & Lag 41 & -0.80 & 15.58 & 16 & -2.00 & 3.00 &  $610 \pm 111$ \\
 &  & Lag 32 & -0.80 & 15.58 & 2 & -0.80 & 0.80 &  $113 \pm 30$ \\
 &  & Lag 42 & -0.80 & 15.58 & 8 & -1.80 & 2.80 &  $444 \pm 107$ \\
 &  & Lag 43 & -0.80 & 15.58 & 8 & -1.80 & 2.80 &  $197 \pm 110$ \\
GRB090424 & 350311 & Lag 21 & -0.94 & 4.95 & 1 & -0.10 & 0.25 &  $20 \pm 12$ \\
 &  & Lag 31 & -0.94 & 4.95 & 2 & -0.10 & 0.25 &  $29 \pm 13$ \\
 &  & Lag 41 & -0.94 & 4.95 & 4 & -0.10 & 0.25 &  $39 \pm 15$ \\
 &  & Lag 32 & -0.94 & 4.95 & 1 & -0.10 & 0.25 &  $23 \pm 9$ \\
 &  & Lag 42 & -0.94 & 4.95 & 4 & -0.10 & 0.25 &  $27 \pm 13$ \\
 &  & Lag 43 & -0.94 & 4.95 & 4 & -0.20 & 0.30 &  $17 \pm 9$ \\
GRB090618 & 355083 & Lag 21 & 46.01 & 135.35 & 4 & -1.00 & 1.00 &  $255 \pm 21$ \\
 &  & Lag 31 & 46.01 & 135.35 & 4 & -1.00 & 1.00 &  $447 \pm 26$ \\
 &  & Lag 41 & 46.01 & 135.35 & 4 & -1.00 & 2.00 &  $894 \pm 43$ \\
 &  & Lag 32 & 46.01 & 135.35 & 4 & -1.00 & 1.00 &  $173 \pm 18$ \\
 &  & Lag 42 & 46.01 & 135.35 & 4 & -1.00 & 1.50 &  $483 \pm 34$ \\
 &  & Lag 43 & 46.01 & 135.35 & 4 & -1.00 & 1.50 &  $283 \pm 34$ \\
GRB090715B & 357512 & Lag 21 & -4.80 & 21.06 & 16 & -2.50 & 2.50 &  $288 \pm 117$ \\
 &  & Lag 31 & -4.80 & 21.06 & 16 & -2.50 & 2.50 &  $732 \pm 127$ \\
 &  & Lag 41 & -4.80 & 21.06 & 64 & -2.50 & 3.00 &  $1080 \pm 224$ \\
 &  & Lag 32 & -4.80 & 21.06 & 8 & -2.50 & 2.50 &  $470 \pm 100$ \\
 &  & Lag 42 & -4.80 & 21.06 & 32 & -2.50 & 2.50 &  $928 \pm 229$ \\
 &  & Lag 43 & -4.80 & 21.06 & 32 & -2.50 & 2.50 &  $375 \pm 215$ \\

\enddata
\end{deluxetable}

\begin{deluxetable}{cccc}
\tabletypesize{\scriptsize} \tablecaption{Correlation coefficients
and fit parameters\label{tab:eband}} \tablewidth{0pt}
\tablehead{\colhead{Channels} & \colhead{Correlation} &
\colhead{Best Fit} & \colhead{$\chi^2$/ndf} }

\startdata

Channel 21 & -0.63 & $\log L_{\rm iso} = (54.8 \pm 0.2) - (1.4 \pm 0.1) \log Lag 21 (1+z)^{-1}$ & 189.4/19 \\
Channel 32 & -0.66 & $\log L_{\rm iso} = (54.5 \pm 0.2) - (1.2 \pm 0.1) \log Lag 32 (1+z)^{-1}$ & 216/25\\
Channel 31 & -0.60 & $\log L_{\rm iso} = (55.5 \pm 0.2) - (1.5 \pm 0.1) \log Lag 31 (1+z)^{-1}$ & 410.8/26\\
Channel 43 & -0.77 & $\log L_{\rm iso} = (55.0 \pm 0.3) - (1.4 \pm 0.1) \log Lag 43 (1+z)^{-1}$ & 109/20\\
Channel 42 & -0.75 & $\log L_{\rm iso} = (55.4 \pm 0.1) - (1.4 \pm 0.1) \log Lag 42 (1+z)^{-1}$ & 178.8/23\\
Channel 41 & -0.67 & $\log L_{\rm iso} = (56.7 \pm 0.3) - (1.8 \pm 0.1) \log Lag 41 (1+z)^{-1}$ & 212.1/25\\

\enddata

\end{deluxetable}

\begin{deluxetable}{ccccc}
\tabletypesize{\scriptsize} \tablecaption{Correlation coefficients
with various corrections\label{tab:corr_correction}}
\tablewidth{0pt} \tablehead{\colhead{Channels} & \colhead{No
correction} & \colhead{z-correction}  & \colhead{k-correction} &
\colhead{both corrections}} \startdata

Channel 21 & -0.38 & -0.63 & -0.29 &  -0.55 \\
Channel 32 & -0.43 & -0.66 & -0.33 &  -0.60 \\
Channel 31 & -0.39 & -0.60 & -0.31 &  -0.54 \\
Channel 43 & -0.61 & -0.77 & -0.54 &  -0.73 \\
Channel 42 & -0.58 & -0.75 & -0.51 &  -0.71 \\
Channel 41 & -0.43 & -0.67 & -0.32 &  -0.61 \\

\enddata

\end{deluxetable}








\end{document}